# Cosmic Temperature Fluctuations from Two Years of *COBE*[1] DMR Observations

C. L. Bennett[2], A. Kogut[3], G. Hinshaw[3], A. J. Banday[4], E. L. Wright[5], K. M. Górski[4,6], D. T. Wilkinson[7], R. Weiss[8], G. F. Smoot[9], S. S. Meyer[10], J. C. Mather[2], P. Lubin[11], K. Loewenstein[3], C. Lineweaver[9], P. Keegstra[3], E. Kaita[3], P. D. Jackson[3], and E. S. Cheng[2]

To Appear in The Astrophysical Journal

---

[2]Code 685, NASA Goddard Space Flight Center, Greenbelt, MD 20771, E-mail: bennett@stars.gsfc.nasa.gov

[3]Code 685.9, NASA Goddard Space Flight Center, Greenbelt, MD 20771

[4]Universities Space Research Association, Code 685.9, NASA/GSFC, Greenbelt, MD 20771

[5]University of California at Los Angeles, Astronomy Department, Los Angeles, CA 90024-1562

[6]On leave from Warsaw University Observatory, Poland.

[7]Princeton University, Department of Physics, Princeton, NJ 08540

[8]MIT, Department of Physics, Room 20F-001, Cambridge MA 02139

[9]Lawrence Berkeley Laboratory, Bldg 50-351, University of California at Berkeley, Berkeley, CA 94720

[10]Enrico Fermi Inst. & Dept. of Astronomy & Astrophysics, Univ. of Chicago, Chicago, IL 60637

[11]University of California at Santa Barbara, Physics Department, Santa Barbara, CA 93106

## ABSTRACT

The first two years of *COBE* Differential Microwave Radiometers (DMR) observations of the cosmic microwave background (CMB) anisotropy are analyzed and compared with our previously published first year results. The results are consistent, but the addition of the second year of data increases the precision and accuracy of the detected CMB temperature fluctuations. The two-year 53 GHz data are characterized by RMS temperature fluctuations of $(\Delta T)_{rms}(7°) = 44 \pm 7$ $\mu$K and $(\Delta T)_{rms}(10°) = 30.5 \pm 2.7$ $\mu$K at 7° and 10° angular resolution respectively. The $53 \times 90$ GHz cross-correlation amplitude at zero lag is $C(0)^{1/2} = 36 \pm 5$ $\mu$K (68% CL) for the unsmoothed (7° resolution) DMR data. A likelihood analysis of the cross correlation function, including the quadrupole anisotropy, gives a most likely quadrupole-normalized amplitude, $Q_{rms-PS}$, of $12.4^{+5.2}_{-3.3}$ $\mu$K (68% CL) and a spectral index $n = 1.59^{+0.49}_{-0.55}$ (68% CL) for a power law model of initial density fluctuations, $P(k) \propto k^n$. With $n$ fixed to 1.0 the most likely amplitude is $17.4 \pm 1.5$ $\mu$K (68% CL). Excluding the quadrupole anisotropy we find $Q_{rms-PS} = 16.0^{+7.5}_{-5.2}$ $\mu$K (68% CL), $n = 1.21^{+0.60}_{-0.55}$ (68% CL), and, with $n$ fixed to 1.0 the most likely amplitude is $18.2 \pm 1.6$ $\mu$K (68% CL). Monte Carlo simulations indicate that these derived estimates of $n$ may be biased by $\sim +0.3$ (with the observed low value of the quadrupole included in the analysis) and $\sim +0.1$ (with the quadrupole excluded). Thus the most likely bias-corrected estimate of $n$ is between 1.1 and 1.3. Our best estimate of the dipole from the two-year DMR data is $3.363 \pm 0.024$ mK towards Galactic coordinates $(\ell, b) = (264.4° \pm 0.2°, +48.1° \pm 0.4°)$, and our best estimate of the RMS quadrupole amplitude in our sky is $6 \pm 3$ $\mu$K (68% CL).

*Subject headings:* cosmology: cosmic microwave background - large scale structure of the universe - observations

## 1. Introduction

The purpose of the Differential Microwave Radiometers (DMR) experiment on the *Cosmic Background Explorer (COBE)* satellite is to measure the large angular scale anisotropy of the cosmic microwave background (CMB) radiation by mapping the temperature of the entire microwave sky at three wavelengths. In this paper, we assume that the CMB is the remnant afterglow from a hot, dense, early universe. The temperature anisotropy on large angular scales reflects the gravitational potential fluctuations at the epoch of last scattering of the CMB photons (Sachs & Wolfe 1967) about 300,000 years after the big bang. Although the big bang model is strongly rooted in the secure observations of the expansion of the universe (Hubble 1929; Jacoby et al. 1992), the abundance ratios of the light elements (Alpher, Bethe & Gamow 1948; Peebles 1966; Walker et al. 1991), and the existence of the blackbody CMB radiation (Penzias & Wilson 1965; Dicke et al. 1965; Mather et al. 1994), it makes no specific prediction of the level of CMB anisotropy. Years of unsuccessful searches for fluctuations in the CMB temperature placed increasingly severe upper limits on the fluctuation amplitude (see, e.g., Wilkinson 1987 for a review).

The lack of detectable large angular scale fluctuations was difficult to explain since regions of the sky separated by more than a couple of degrees were never in causal contact in the history of the universe, and thus had no way to establish a uniform temperature with such high precision: the *horizon problem* (Weinberg 1972; Misner, Thorne, & Wheeler 1973). The simple big bang model takes the isotropy of the CMB as an initial condition. Also, measurements have long indicated that our local universe is nearly flat. To account for this in the big bang model the flatness must be set as an initial condition with extreme accuracy: the so-called *flatness problem* (Dicke & Peebles 1979). The inflation scenario (Guth 1981) describes a phase transition in the early universe that drives an exponential expansion of the universe. In this way the horizon problem is alleviated, since our entire observable universe inflated from a small region that was in causal contact at an early epoch, and the flatness problem is reduced since inflation drives the spatial curvature radius to infinity. In principle a full theory of the early universe could predict the amplitude and spectrum of CMB temperature fluctuations, but there currently is no such generally accepted theory.

The *COBE* DMR experiment was designed to detect and characterize the CMB anisotropy. Smoot et al. (1990) and Bennett et al. (1992a) provide detailed descriptions of the DMR experiment. Smoot et al. (1992), Bennett et al. (1992b), Wright et al. (1992), and Kogut et al. (1992) reported the detection of CMB temperature fluctuations based on the first year of DMR data. Bennett et al. (1992a) presented the DMR calibration and its uncertainties and Kogut et al. (1992) gave a detailed treatment of the upper limits on residual systematic errors affecting the first year of data. Bennett et al. (1992b) showed that spatially correlated Galactic free-free and dust emission could not mimic the frequency spectrum nor the spatial distribution of the observed fluctuations. Bennett et al. (1993) show that the pattern of fluctuations does not spatially correlate with known extragalactic source distributions. Confirmation of the *COBE* results was attained by the positive cross-correlation between the *COBE* data and data from balloon-borne observations at a shorter wavelength (Ganga et al. 1993). The proper motion of our solar system, relative to the Hubble expansion, gives rise to a dipole anisotropy of the CMB, first reported by Conklin (1969) and Henry (1971). Kogut et al. (1993) presented the best estimate of the dipole parameters from the *COBE* DMR based on the first year of data, and Fixsen et al. (1993) presented the consistent *COBE* Far Infrared Absolute Spectrophotometer (FIRAS) dipole results.

Since fluctuations in the gravitational potential cause both temperature anisotropy in the CMB and the gravitational clustering of mass in the universe, a cosmological model must simultaneously account for both. The pattern of CMB fluctuations was predicted by Peebles & Yu (1970), Harrison (1970), and Zel'dovich (1972) to be scale invariant, with equal RMS gravitational potential fluctuations on all scales. A scale invariant spectrum is also a natural consequence of the inflationary model. The *COBE* measurements are consistent with such a spectrum. Wright et al. (1992) compared the amplitude of CMB anisotropy with large scale galaxy clustering within the context of an array of theoretical models from Holtzman (1989). Wright et al. found that standard cold dark matter is somewhat deviant, and that a model with hot plus cold dark matter (i.e. mixed dark matter) and a model with a cosmological constant plus cold dark matter fit the data well. Some detailed examinations of the mixed dark matter model continue to show that it is in excellent agreement with most data (Schaefer & Shafi 1993; Holtzman & Primack 1993; Klypin et al. 1993; Fisher et al. 1993; Davis, Summers, & Schlegel 1992) while others (e.g. Cen & Ostriker 1993; Baugh & Efstathiou 1993) report problems with the model. Cold dark matter models with a non-zero cosmological constant have been considered in detail by Stompor & Górski (1993), Efstathiou, Bond, & White (1992), Cen, Gnedin, & Ostriker (1993), and Kofman, Gnedin, & Bahcall (1993). The possibility of a cold dark matter universe, but with a slightly tilted spectrum of spatial fluctuations has been considered by Adams et al. (1992), Cen et al. (1992), Liddle, Lyth, & Sutherland (1992), Lidsey & Coles (1992), and Lucchin, Matarrese, & Mollerach (1992). In open universe models, examined by Wright et al. and in more detail by Kamionkowski & Spergel (1993), the fluctuations observed by *COBE* may arise from the decay of potential fluctuations at redshifts, $z < \Omega^{-1}$ rather than from the Sachs-Wolfe potential fluctuations at the last scattering surface at $z \sim 1000$. Rather than postulating primordial curvature fluctuations to seed the gravitational formation of structure in the universe, models with topological defects such as cosmic strings (Vilenkin 1985, Stebbins et al. 1987), global monopoles (Bennett & Rhie 1993), cosmic textures (Gooding et al. 1992), or domain walls (Stebbins & Turner 1989; Turner, Watkins, & Widrow 1991) have been considered. Some have suggested that our understanding of the laws of gravity fail on large scales, and that an alternate gravity theory is needed (e.g. Mannheim & Kazanas 1989; Mannheim 1992, 1993).

Despite the many successes of cosmology, the large scale structure problem remains to be definitively solved and the theoretical scenarios must be constrained by observations. The *COBE* DMR experiment has taken data on large angular scale ($> 7°$) CMB anisotropy for four years and this paper presents an analysis of the first two years of data, from 1989 December 22 to 1991 December 21. The DMR experiment consists of six differential microwave radiometers; two nearly independent channels (A and B) at each of three frequencies: 31.5, 53, and 90 GHz (wavelengths 9.5, 5.7, and 3.3 mm). Each radiometer measures the difference in power, expressed as a differential antenna temperature, between two $7°$ regions of the sky separated by $60°$. The combined motions of spacecraft spin (73 s period), orbit (103 min period) and orbital precession ($\sim 1°$/day) allow each sky position to be compared to all others through a highly redundant set of all possible difference measurements spaced $60°$ apart (Boggess et al. 1992). The full sky is sampled every six months.

The DMR has three separate receiver boxes, one for each frequency, that are mounted $120°$ apart on the outside of the cryostat containing the FIRAS and the Diffuse Infrared Background Experiment (DIRBE). The DMR has ten horn antennas, all of which have an approximately Gaussian main beam with a $7°$ full width at half maximum (FWHM) (see Wright et al. (1994a) for a more detailed description of the DMR beam pattern). The pair of horns for each channel are pointed $60°$ apart, $30°$ to each side of the spacecraft spin axis and are designated as Horn 1 and Horn 2. The differential temperature is measured in the sense Horn 1 minus Horn 2. For each channel, the switching between Horns 1 and 2 is at a rate of 100 Hz, and the switched signals undergo amplification, detection, and synchronous demodulation with a 0.5 s integration period. The 53 and 90 GHz channels use two separate linearly polarized horn pairs, while the 31 GHz channels receive opposite circular polarizations in a single pair of horns. The E-planes of linear polarization for all 53 and 90 GHz channels are directed radially outward from the spacecraft spin axis. A shield surrounds the aperture plane and shields all three instruments from solar and terrestrial emission. See Figure 8 of Bennett al. (1992a) for a schematic of the *COBE* aperture plane.

In §2 of this paper we describe the software system and data processing of the first two years of *COBE* DMR data. In §3 we discuss the calibration and in §4 we discuss the systematic error analysis. In §5 we present the basic scientific results and in §6 we summarize. Separate papers present additional analyses of the two-year DMR anisotropy results in terms of modified spherical harmonic coefficients on the sphere with a Galactic plane cut (Wright et al. 1994b), and in terms of newly-defined orthogonal functions on the cut sphere (Górski 1994; Górski et al. 1994).

## 2. The Software System & Data Processing

The purpose of the DMR software system is to take the raw telemetry data from the instrument and produce calibrated maps of the sky that have instrumental and environmental systematic effects reduced below specified levels. The techniques used and resulting systematic error limits for the first year of DMR data are described by Kogut et al. (1992). A description of the lower systematic error limits that apply to the two years of data are reported in this paper. The software system and data processing algorithms are also described by Janssen & Gulkis (1991) and Jackson et al. (1992).

Having gained experience in processing the first year of *COBE* DMR data we implemented many changes in the software for the analysis of the first two years of data. Many of these changes were made to reduce the size of intermediate files and increase the speed and efficiency of the data processing. Other changes were made to improve the data quality for scientific analysis.

The DMR telemetry consists of uncalibrated differential temperatures, taken every 0.5 seconds for each of the six channels, plus housekeeping data (voltages, temperatures, relay states, etc.). These are merged with spacecraft attitude data into a time-ordered data set. Various algorithms are used to check data quality and flag bad data (Keegstra et al. 1992).

Our previous analysis of the first year of DMR data excluded data taken when the Moon was within 25° of an antenna beam center. This proved to be overly conservative, so in the current analysis we change to a 21° lunar cut, gaining 20% more data in the sky map pixels near the ecliptic plane. In the current analysis 4.5% of the data are rejected because the Moon was within 21° of an antenna beam center.

We have now included corrections for the planets Mars and Saturn in addition to Jupiter, the only planet for which corrections were applied previously. Mercury and Venus are never near the beam centers and the contributions from Uranus, Neptune, and Pluto are negligible compared with other systematic errors.

As in our earlier analysis, we flag as bad all data when the limb of the Earth was higher than 1° below the plane of the shield (3° at 31 GHz). This excludes 6% of the DMR data.

Near the times of the summer solstice, the *COBE* spacecraft enters into the Earth's shadow during a portion of each orbit and is thus 'eclipsed'. This causes changes in the spacecraft temperature and electrical systems. Systematic errors during the eclipse season significantly affect the DMR 31 GHz channels. Hence, the data reported in this paper do not include 31 GHz data for the period 0 UT on 1990 May 21 through 0 UT on 1990 July 25; and from 0 UT on 1991 May 21 through 0 UT on 1991 July 25.

On 1991 October 4, the 31B channel suffered a permanent increase in receiver noise by more than a factor of two for unknown reasons. Data following this event are not included in the analyzed data set reported in this paper.

We remove the instrumental offset by fitting a smooth baseline to the uncalibrated time-ordered data. As a cross-check, baselines were routinely fitted three different ways: a running mean, a slowly varying cubic spline, and a more rapidly varying cubic spline fit to the data after correction for magnetic susceptibility (to be discussed in §4.1). The averaging period for the running mean is two orbit periods (the orbit period is 103 min); the slowly varying spline can follow variations in the baseline that occur more slowly than every 17 min; and the fast spline can follow variations that last three minutes or more. The three baselines represent a trade-off between fitting systematic errors using a priori functional forms versus removing them as part of the baseline. Results discussed in this paper use the running mean baseline.

Calibration of the data to antenna temperature is achieved by turning on noise sources of known antenna temperature for two minutes every two hours. Long-term stability of the noise sources is checked by observations of the Moon and by the amplitude of the modulation of the CMB signal over the course of a year caused by the Earth's 30 km s$^{-1}$ orbital motion. See Bennett et al. (1992a) for a detailed description of DMR calibration techniques, which are applied in §3, below.

We correct the calibrated time-ordered data for known instrumental and systematic effects. We correct for the Doppler effect from the *COBE* velocity about the Earth ("satellite velocity") and the Earth's velocity about the solar system barycenter ("Earth velocity"); the susceptibility of the instrument to the Earth's magnetic field; emission from the Moon in the antenna sidelobes; the slight 'memory' of the previous observation (arising in the lock-in amplifier); and emission from the planets Jupiter, Mars, and Saturn.

In processing the first year of data we set a flag to exclude data within 10° of the Galactic plane for the determination of several important parameters. Since then we have found that some Galactic effects remain in the estimation of the calibration and other systematic errors so the flag has been increased to 15° for these estimates.

In the earlier first year data analysis Kogut et al. (1992) reported that the DMR lock-in amplifier (LIA) retained some memory of the previous observations. This memory effect was taken into account in the systematic error analysis of Kogut et al. We caution that the first nonzero bin of the two-point correlation function in our first year results paper was excluded because of this effect. In the current two year analysis we correct for this effect in the software, making the first nonzero correlation bin useful; thus we have retained it.

The corrected, calibrated time-ordered data were sorted and combined according to the sky pixels seen by Horns 1 and 2. (Note that linear polarization information for the 53 and 90 GHz data is retained since the E-plane is perpendicular to the great circle joining the two pixels.) For each channel, there are over 1,600,000 permutations of the 6144 pixels, most of which were sampled in the first six months of the mission (after six months the number of newly sampled permutations grows slowly since *COBE*'s orbital plane has completed its initial sweep of the sky). Thus we have a highly redundant set of data in the form of temperature differences between pairs of pixels. We form the $\chi^2$ sum of the measurements involving each pixel, and force derivatives with respect to the pixel temperatures to be zero. There are P normal equations in the pixel temperatures, where P is the number of pixels. Because of the 60° constraint of the horn separations, the P × P matrix of normal equation coefficients is only a few percent filled; it is a sparse matrix. We solve for the sky temperature of each pixel simultaneously with coefficients of systematic error models using iterative techniques (Janssen & Gulkis 1992). Since the instrument only measures temperature differences between different sky directions, these maps of the sky only represent differences of the sky temperature (the DMR differential data would be identical if the whole sky changed in temperature by a constant amount).

The effects of errors inherent in pixelizing over the non-uniformly sampled sky are largest for the two strongest signals in the DMR maps: the dipole signal and the Galactic signal. To reduce the error we corrected the calibrated time-ordered data, before pixelizing, for a nominal CMB dipole of 3.325 mK (in thermodynamic temperature) directed towards J2000 R.A. = $11^h$ $12^m$ $57^s.6$, Dec = $-6° 0' 36''$. The time-ordered data are pixelized according to the quadrilateralized spherical cube projection (White & Stemwedel 1992), which projects the entire sky onto six cube faces. Each face is pixelized into $2^{2(N-1)}$ approximately equal-area, square pixels where $(N-1)$ is called the index-level. Our previous data processing created maps of the sky with fixed $(N = 6)$ 2.6° pixels (i.e. 6144 pixels, each subtending 6.7 square degrees), but in the current analysis we make split resolution maps, with $(N = 6)$ 2.6° pixels for Galactic latitudes $|b| > 20°$ and $(N = 7)$ 1.3° (1.7 square degree) pixels for Galactic latitudes $|b| < 20°$. This allows somewhat more resolution of the Galactic plane to minimize the effects of gradients in the Galactic disk emission. After obtaining the final map solutions, we average the higher resolution pixels together to produce sky maps where all of the pixels are 6.7 square degrees.

The noise level of the sky maps varies by more than a factor of two over the sky owing to differences in sky coverage during the mission. The greatest redundancy is on rings of 60° diameter approximately centered at the North and South ecliptic poles since those regions are sampled on every orbit. The least redundancy is near the ecliptic plane owing to the presence of the Moon. For channels 31A and 31B, the more stringent limits on the position of the limb of the Earth relative to the shield, and the rejection of data during the two eclipse months also reduces coverage considerably for certain pixels. Table 1 gives the maximum, mean, and minimum coverages and the corresponding pixel-to-pixel RMS noise levels.

## 3. Calibration

### 3.1. Calibration Techniques

In this section we follow the DMR calibration techniques described by Bennett et al. (1992a) and update the results to cover the first two years of *COBE* DMR data. The primary calibration for the DMR instrument is based on the on-board noise sources, which regularly inject small signals into the amplification chain. Uncertainties or errors in the primary calibration affect the DMR analysis in three ways: (1) The magnitude of detected structure is only known to the precision of the absolute calibration, (2) Errors in the absolute calibration create artifacts in maps from which astrophysical emission (e.g., Galactic emission or the Earth Doppler) has been removed, and (3) Drifts in the calibration create artifacts in the map because different parts of the sky were observed with different calibrations.

The noise sources are both observed in each DMR channel. The ratio of the two noise source signals in each channel provides a lower limit to the stability of the noise-source derived calibration. Secondary calibration signals are available from the Moon, the Doppler effect of the Earth and satellite velocity about the solar system barycenter, the cosmic dipole, and the instrument total power. Comparison of these secondary calibrators to the primary calibration provides limits to the absolute calibration accuracy and the magnitude of any drifts with time.

The calibration has been examined for the first two years of DMR data, excluding data from the 31A and 31B channels during the eclipse seasons, and data from the 31B channel when its noise increased. The pre-flight calibration has been adjusted for four effects: (1) a 31B gain decrease of 4.9%, (2) a 90B gain increase of 1.4%, (3) a 0.7% step in the 90 GHz noise source B emitted power, and (4) a linear drift in both 90 GHz gain solutions (0.80 and 0.92 % yr$^{-1}$). Effects (1) and (2) are adjustments to our estimated values from pre-flight to in-flight, while (3) and (4) are real observed changes seen in-flight. With these corrections, the absolute and relative calibrations inferred from the Moon, Earth Doppler dipole, and cosmic dipole all agree within the precision of the (adjusted) ground results. Current uncertainties in the absolute calibration and in linear calibration drifts are given in Table 2, and are discussed below. The first column specifies the radiometer channel and the second column gives the corrections to, and the uncertainties of, the noise source calibration. We used the noise source calibration determined before launch as our primary in-flight calibration so the mean 'Ground' corrections in the Table are necessarily zero with ground calibration uncertainties as indicated. The third column, 'Flight', gives the ratio of either the lunar-derived or Doppler-derived absolute calibration (whichever was more sensitive) to the noise source-derived calibration. The flight corrections are not statistically significant, so we do not apply them. The fourth column gives corrections to the gain ratios (channel A/channel B) determined from observations of the Moon. The bounds on linear calibration drifts, and on orbital and spin period modulations of the calibration, are given in the fifth through seventh columns. We conclude that the absolute calibration results for the first two years of data are accurate to within the uncertainties of the ground-based calibration, and that drifts in the calibration are small.

### 3.2. Absolute Calibration

We use DMR observations of the 0.3 mK dipole from the Earth's velocity about the barycenter of the solar system to derive an independent estimate of the DMR absolute calibration. We derive a gain correction factor deduced from the Earth velocity dipole:

$$\text{Dipole in data} = A \times \text{Predicted Earth dipole} \tag{1}$$

where the "predicted" dipole is given by the usual Doppler formula for a 2.73 K blackbody CMB (Mather et al. 1990, 1994) and the Earth's known velocity vector. The calibration factor, $A$, can be derived both from the time-ordered-data and from the sparse matrix equation. The Earth Doppler pattern is not orthogonal to other structure in the sky, but this cross talk should be minimized in the sparse matrix solution, which solves for the fixed sky temperature in 6144 pixels simultaneously with the moving Earth Doppler dipole. Results are given in column 3 of Table 3. The uncertainties in the Doppler absolute calibration are a factor two to four larger than the uncertainties in the ground calibration. The Doppler calibration results provide marginal evidence of ground calibration errors, but the deviations are sufficiently small that any corrections would have a negligible effect on the final anisotropy determination.

DMR observations of the Moon allow an independent calibration. We derive the gain by comparing the signal change, in the telemetry digital units (du), caused by the Moon with a model of the antenna temperature expected from the Moon's position in the beam and lunar emission properties (Keihm 1982, 1983; Keihm & Gary 1979; Keihm & Langseth 1975). The gain derived from the Moon signal is observed to depend on both lunar phase and time of year, which we take as an indication that the model is not perfect. The lunar gain errors have peak-to-peak amplitudes of 3.7, 5.3, and 6.2 % as a function of the lunar phase for 31.5, 53, and 90 GHz, respectively, and there is a 2% peak-to-peak gain variation as a function of the time of year for all three frequencies. We attribute these to inaccuracies in the model of lunar microwave emission and not to software errors in our implementation of this model. The annual gain is fit to a sinusoid and the monthly phase variation to a spline to empirically remove these periodic effects from the lunar gains. The resulting set of "stable" lunar gains can be compared to the noise-source gain solutions and are given in Table 3. The error bars are 68% confidence statistical uncertainties from the lunar and noise source uncertainties and include the 6% systematic uncertainty evident in the phase- and time-dependent periodic variations in the lunar gain. The mean of all six channels shows a lunar gain 1.84% larger than the noise source solution, well within the phase and annual uncertainties of the lunar model.

The lunar-derived gains also allow a check of the relative calibration of the A and B channels at each frequency. Since the systematic uncertainties in the lunar model, including its phase and annual effects, should not depend on channel, the ratio of lunar-derived to noise source-derived gains in the A and B channels should reflect the relative error in the noise source gain solution. Results are given in Table 4; the noise source-derived gains have been adjusted, as described in §3.1. There is general agreement between the Earth Doppler, lunar, and noise source methods. Although there is evidence for relative calibration errors, they are at a level less than 0.4%, well below the overall level of uncertainty for the absolute calibration (see Table 3). There is no evidence for absolute calibration errors at the 95% confidence level.

### 3.3. Calibration Drifts

In this section we consider limits that can be placed on calibration changes with time using four techniques: changes in the 3 mK dipole amplitude with time, changes in the Moon to noise source signal ratio with time, changes in the data RMS with time, and changes in the ratios of noise source amplitudes with time.

We use observations of the 3 mK dipole to limit the amplitudes of drifts in the noise source gain solution. Drifts in the gain solution will create an apparent time-dependent change in the dipole amplitude. We examine the dipole amplitude in the time-ordered-data by correcting the data for our best estimate of systematic errors, and then by fitting a linear function of time to the amplitudes of a dipole plus quadrupole signal. The results, expressed as a percentage calibration drift per year, are given in Table 5. Since the calibrated dipole amplitude is given by

$$\text{Calibrated dipole} = \text{True dipole} \times (\text{true gain}/\text{ noise source gain}) \tag{2}$$

a positive drift in the apparent dipole amplitude corresponds to an increase in the true instrument gain relative to the noise source gain solution. Using a similar analysis we read the time-ordered data and fit directly to a function of the form

$$DT = [1 + B(t)] \times \text{Dipole}(X, Y, Z) \tag{3}$$

That is, we perform a nonlinear fit to a dipole of fixed amplitude and direction as well as a linear fractional gain drift. No channel shows a significant drift in the dipole amplitude.

The lunar-derived gain solutions, corrected for the periodic phase and annual variations discussed above, serve as a stable external reference to the internal noise sources. The lunar-to noise source-derived gain ratios are fit for a linear drift in time (Table 5). Only the 90A and 90B channels show a significant drift at the 95% confidence level. Recall that we have already corrected for drifts of 0.80% $\text{yr}^{-1}$ for 90A and 0.92% $\text{yr}^{-1}$ for 90B based on prior analyses of the data (see §3.1). We now find residual gain drifts of 0.15% $\text{yr}^{-1}$ for 90A and 0.18% $\text{yr}^{-1}$ for 90B. These residual gain drifts are small enough that they create negligible artifacts in the maps and do not require computationally expensive processing of the data. Future proccessing will continue to iterate towards the best gain solutions.

Each noise source is seen in both channels. The ratio of the two noise amplitudes observed in the same channel is used as a diagnostic for changes in noise source antenna temperature. Since the observed noise source amplitude (in du) is the product of the noise source antenna temperature and the gain, the ratio for each channel is independent of the actual channel gain and instead gives the ratio of the noise source antenna temperatures. Changes or drifts in antenna temperature that are not properly accounted for will result in incorrect gains applied to the time-ordered data. The ratio of the A/B noise source amplitudes are binned into weekly values and linear drifts in antenna temperature are fit. The results are given in Table 5. Only the 31A, 31B, and 53B channels are reasonably fit by a simple linear drift for two full years. The 53A channel noise source ratio is constant throughout the first year before beginning a +0.2 % yr$^{-1}$ linear drift in the second year of data. Both 90 GHz noise source ratios vary at the 0.2% level for the first year before beginning a +0.6 % yr$^{-1}$ linear drift in the second year. The failure of the A/B noise source ratios to agree between the two channels indicates that their trends are caused by a process more complicated than a simple change in broadcast power (antenna temperature).

Changes in the instrument calibration at the orbit and spin periods (caused, for instance, by differential solar heating) can create artifacts in the DMR maps. We test for these effects by binning the raw noise source amplitudes, the total power, and the instrument RMS at the spin and orbit periods. The total power shows clear structure at the orbit period but it is not clear whether this reflects a true calibration change or a change in system temperature. The data RMS shows large variations as a function of orbit angle, linked to real structure in the sky (the Galaxy) and not to gain variations. Artifacts created by gain variations of this amplitude are negligible (see §4).

### 3.4. Calibration Summary

Table 3 summarizes the absolute calibration of the DMR radiometers. The ground-based calibration remains the most sensitive. Statistical errors in the lunar gain correction are dwarfed by the unexplained systematic changes with lunar phase and time of year. By taking the ratio of the A/B channels we can cancel systematics common to both channels and probe the absolute calibration more sensitively. As seen in Table 4, relative calibration differences between the A and B channels, while statistically significant, are within 0.2% and thus serve as a useful cross-check on the individual A and B channel absolute calibrations. Calibration errors at this level produce negligible effects in the final maps. Table 5 presents a summary of the gain drifts (true gain/noise source gain) from each technique. The drifts inferred from the dipole, Moon, and noise source ratios are in general agreement and provide some evidence that calibration drifts are dominated by changes in the power emitted by the noise sources. The drifts inferred from the data RMS are inconsistent with a simple gain drift and are dominated by changes in the receiver system temperature. The Moon is the brightest stable external reference, and provides the best limits to time-dependent errors in the noise-source gain solution. There may be gain drifts correlated with orbit angle, but the systematic effects of these are limited (see §4.3).

## 4. Systematic Errors

Systematic effects in the DMR time-ordered differential data can produce large-scale artifacts in the maps. We create sky maps of systematic effects using attitude information and specified models of the systematic signals as a function of time as inputs. The overall amplitude of the resultant map is proportional to the amplitude of the signal in the time domain, while the pattern is determined by the details of the signal and the sky coverage. We assume that uncertainties in the systematic error maps are dominated by amplitude uncertainties and we neglect any changes in the patterns themselves that would be caused by deviations from the assumed time dependence or sky coverage. Most sources of uncertainty that are external to DMR (e.g., uncertainty in the microwave emission from the Moon or the Earth) are identical in all six channels and will cancel when two channels are differenced. A fraction of the total uncertainty (e.g., radiometer calibration) is channel-specific and will not cancel in the difference between two channels.

Uncertainties are expressed as a fraction of the amplitude of each systematic effect. The uncertainty map is found by multiplying the map of each effect, if uncorrected, by the fractional uncertainty in the correction. Tables 11 through 16 summarize our best estimates of the amplitudes of systematic effects in the DMR sky maps before corrections are applied, and 95% confidence upper limits on systematic effects after we have applied our best corrections. For each systematic effect we give error estimates in terms of peak-to-peak, RMS, and multipole amplitudes where $\Delta T_\ell$ is the amplitude of the $\ell^{\text{th}}$ spherical harmonic,

$$\Delta T_\ell^2 = \sum_m \frac{|a_{\ell m}|^2}{4\pi},\tag{4}$$

where $a_{\ell m}$ are the spherical harmonic coefficients, $T(\theta, \phi) = \sum_\ell \sum_m a_{\ell m} Y_{\ell m}(\theta, \phi)$, and $\Delta T_2$ is the familiar RMS quadrupole, $Q_{rms}$. In Tables 11 through 16, the rows $\beta_X$, $\beta_R$, and $\beta_T$ refer to magnetic effects on the DMR instrument as the spacecraft travels through the Earth's magnetic field. These are projected onto three coordinate axes, as discussed below. The rows "Earth" and "Moon" refer to estimates of the degree to which microwave emission from the Earth and Moon contaminate the DMR data. The "Doppler" row refers to systematic errors that result from uncertainties in the absolute CMB temperature, Earth and spacecraft velocities, and the radiometer's absolute calibration, as discussed in §3, above. The "Spin" row refers to systematic effects at the *COBE* spin period, discussed below. All other systematic errors are estimated with their amplitudes added together in quadrature, and reported in the row "Other." There are over a dozen effects lumped into "Other," such as thermal susceptibility, radio frequency interference, pointing errors, and numerical errors. For a full listing and discussion of these effects see Kogut et al. (1992). Tables 17 through 22 provide a breakdown of the systematic errors on the quadrupole, $\Delta T_2$.

### 4.1. Magnetic susceptibility

The radiometers are switched at 100 Hz between two horns using a latching magnetic ferrite circulator as a microwave switch. While this technique has the advantages of no moving parts, rapid transition, low insertion loss, and high off-port isolation, it has the major disadvantage of sensitivity to variations in the ambient magnetic field due to the motions of the spacecraft.

We model and remove the magnetic susceptibility effect by including appropriate systematic error fitting terms in the sparse matrix solution. These terms are parameterized by three magnetic susceptibility coefficients, $\beta_X$, $\beta_R$, and $\beta_T$, in mK per Gauss as defined by Kogut et al. (1992). The coordinate system is cylindrical where the $X$ subscript in the coefficient refers to the direction antiparallel to the spacecraft spin axis vector, the $R$ subscript to the outward radial direction, and the $T$ subscript refers to the direction tangential to the spin (defined from horn 1 to horn 2). Magnetic effects along the $X$ axis produce a small dipole in the sky map. In this case the signal is nearly identical in both horns so the differential signal, and hence the effect on the map, is diluted by the ratio of spin period to orbital period. Since the $X$ component varies slowly, some of the $\beta_X$ susceptibility can be removed as part of the baseline and couples into the map solutions only weakly. For example, simulations show that the spline baseline removes about 60% of the $\beta_X$ magnetic effect before the sparse matrix fitting routine gets a chance. Magnetic effects along the $R$ axis produce a large quadrupole. The effect occurs synchronous with the spin rate but at right angles to the antenna pointing. Magnetic effects along the $T$ axis produce a large dipole. The effect occurs synchronous with the spin rate and in the same direction as the antenna pointing, producing a dipole aligned with the Earth's field. There will also be a quadrupole component comparable in amplitude to the $R$ axis quadrupole.

The inclination of the *COBE* orbit with respect to the Earth's field allows separation of magnetic and sky signals over extended observations (few months). Using the sparse matrix of data and functional forms for systematic errors we perform a simultaneous least-squares fit to a sky map and magnetic susceptibility. We fit for a time-independent linear magnetic coupling coefficient to the external magnetic field intensity. The fitted magnetic susceptibilities and uncertainties for our current best estimates are given in Table 6, using the mean baseline removal. The uncertainty in the $\beta_R$ and $\beta_T$ components are the important terms for the DMR maps since we remove the fitted effect from the maps. Only the $\beta_R$ and $\beta_T$ coefficients have significant projection onto the maps. The flight and ground susceptibilities show similar trends (large susceptibilities for the 53A $X$ and $T$ axes and 90A $R$ axis), but the flight results are superior. We assume linear magnetic coupling in our systematic analyses, but we have fit more complicated forms, including tensor, and nonlinear couplings. Fits to these more complicated couplings have a poorer $\chi^2$ than the simpler linear model. We conclude that such couplings contribute less than 1% (95% CL) to the observed variance in the DMR maps.

The derivation of the magnetic susceptibility coefficients depend on an assumed model for the Earth's magnetic field. We use the International Geomagnetic Reference Field from Barker et al. (1986) to order $\ell = 8$. The field model extends to $\ell = 10$, so our cut-off at $\ell = 8$ implies that the uncertainty in our application of the field model is $\sim 0.3$ mGauss. We make use of the magnetometers on the *COBE* spacecraft to check consistency with this magnetic field reference. The rms difference in vector magnitude is 10.7 mG averaged over the 2-year mission, and 9.6 mG if the eclipse season is excluded, corresponding to 6.6% and 5.8% uncertainty in the mean field model, respectively. This is near the digitization limit of the magnetometers. We adopt the error to our implementation of the field model as 6.6% for two years of data, including eclipse data, and 5.8% for two years excluding eclipse data. The local magnetic field arising from the spacecraft torquer bars (electromagnets), used for attitude control, is automatically taken into account in these limits.

We generate maps of the individual $X$, $R$, and $T$ magnetic effects by subtracting maps of the sky made with and without accounting for each magnetic axis effect. Our ability to remove the magnetic signal is limited by uncertainty in the fitted coefficients and uncertainty in the external magnetic fields near the DMR ferrite components. The uncertainty in the magnetic correction is then the quadrature sum of the fractional uncertainty in the fitted coefficients and the uncertainty in the local (e.g. torquer bar generated) magnetic field. The resultant uncertainties are dominated by the uncertainties in the fitted coefficients, which are independent from channel to channel.

If the actual susceptibility differs slightly from the fitted value, the residual magnetic signal in the DMR maps will have the same pattern as the uncorrected maps with amplitude reduced by the ratio of the fitted uncertainty to the fitted coefficient. We analyzed the uncertainty maps as though they were mission sky maps, and have derived the limits to magnetic effects in the DMR sky maps (Tables 11 through 16). After correction, the residual magnetic effects contribute $< 5$ $\mu$K to $Q_{rms}$ in the individual maps of channels A or B, to the $(A + B)/2$ (sum) maps, or $(A - B)/2$ (difference) maps.

### 4.2.    Microwave Emission from the Earth

The Earth, as seen by the DMR experiment, is an extended circular source of emission with a radius of $\sim 61°$ and a mean temperature $\sim 285$ K. The DMR experiment design minimizes contamination of the faint cosmic data from the bright Earth signal by the use of horn antennas with good off-axis sidelobe rejection, the use of a reflecting shield between the DMR antennas and the Earth, and the rejection of data when the Earth signal is predicted to be large.

The spacecraft attitude generally keeps the limb of the Earth entirely below the *COBE* Earth/Sun shield so that its emission can affect the DMR only after diffracting over the shield edge. During the "eclipse season," near the June solstice, the Earth limb rises as high as 8° above the top of the shield. We reject all data taken when the Earth limb is 1° below the top of the shield or higher for the 53 and 90 GHz data, and 3° below the top of the shield for the 31 GHz data.

Scalar diffraction theory is used to predict the signal from the Earth based on the nominal flight configuration of the horn apertures and deployed shield. The model is subject to two major sources of uncertainty: the antenna gain at the top edge of the shield dominates the overall amplitude uncertainty for all limb angles, while the detailed shape of the shield and the shield/antenna geometry dominates the diffraction uncertainties (relative signal change as a function of elevation angle). Numerical estimates of these uncertainties, obtained by varying the shield position by 1° (6 cm in height along the spacecraft spin axis), indicate that the model uncertainty is two to five times the nominal amplitude.

We do not know the deployed shield position and geometry precisely enough to correct the DMR data. We derive upper limits to the Earth signal by fitting the model in narrow ranges of elevation angle to the DMR data binned by the position of the Earth in a spacecraft-fixed coordinate system. Table 7 presents limits to the Earth signal in the time-ordered data ($\mu$K antenna temperature). Limits with "Earth Above Shield" refer to data with Earth limb elevation +1° to +6°, while "Earth Below Shield" refers to limb elevation −1° through −6°.

With the Earth above the shield, a fit to the azimuthal variation predicted by the differential antenna beam shows a positive detection in all channels except the 31A, at an amplitude roughly 1/3 of the predicted signal. This falls well within the uncertainty expected from small changes in the deployed shield position with respect to the nominal position. Given the noise levels of the two-year maps, we would not expect to detect the Earth below the shield. A fit with the Earth just below the shield shows no Earth emission at the 30 $\mu$K level (95% CL). This limit is consistent with less sensitive upper limits derived from methods that do not rely on specific models of the Earth emission (Kogut et al. 1992), and is a factor of three more conservative than scaling the detected signal with the Earth above the shield. Table 8 shows the 95% confidence level upper limits to the Earth emission in the time-ordered data derived from this method.

We derive upper limits to the effect of Earth emission in the DMR sky maps by adopting the time-ordered Earth diffraction model (Table 8) and making maps of the sky with and without this model correction. The Earth contributes less than 0.1% to the observed sky variance. An alternative approach is to replace the fitted diffraction model with the $(A + B)/2$ binned Earth data as the model of Earth emission in this procedure. With these data, the Earth contributes less than 0.6% to the observed sky variance.

### 4.3.   Effects at the Spin and Orbit Period

The DMR time-ordered data are binned according to the spin and orbit periods to place limits on systematic effects with these modulations. We have three estimates of signal amplitude at the orbit period: the eclipse limits, the power spectrum of the data at the orbit period, and the peak-peak scatter in the calibrated, corrected data binned at the spacecraft orbit period. Table 10 gives limits from the three effects. Since orbit binning is intrinsically more sensitive than the FFT, we adopt the upper limits from the binned data as the 95% CL upper limit to combined effects at the orbit period.

We have two estimates of (upper limits to) effects at the spin period: the power spectrum and direct spin-binned data. Table 10 gives limits from these effects. The limits for the FFTs are the 95% CL upper limit for all signals near the spin period; the actual amplitude in the frequency bin containing the spin period is about a factor of two lower. The limits for the spin-binned data are the RMS scatter of the data points (which show no evidence of structure).

We use the upper limits from spin-binned data in each channel to limit possible effects near the spin period. Examples of such effects would be thermal gain changes, for which we can independently establish comparable limits. We model this effect using a sparse matrix systematic solution with a sine wave locked to the solar angle.

### 4.4.   Miscellaneous

In Kogut et al.'s (1992) detailed description of the systematic error analysis of the first year DMR data the limits placed on many of the potential systematic errors are severe. We have placed upper limits on all of the same potential systematic errors for this analysis of the two year data, but since several of the limits are small we will not individually discuss them in detail here. We note that systematic error studies include instrument cross-talk, seasonal effects, solution convergence of the sparse matrix, pixel independence, baseline subtraction, the effects of nonuniform sky coverage, the effects of discrete pixelization, artifacts from the instrument, radiation from the *COBE* Sun/Earth shield, radio frequency interference, radiation from the Sun, and residual radiation from the Moon and Planets. Our new combined upper limits on all of these effects are included in our overall stated quantitative errors.

### 4.4.1. Seasonal Effects

The 31 GHz radiometers show significant anomalous behaviour during the eclipse season when sunlight is blocked from *COBE* by the Earth for a portion of each orbit. For this reason we do not use the 31 GHz data from this season. We now also detect a small orbitally modulated signal during the eclipse season in the 53 and 90 GHz radiometer data as well. The 31 GHz A-channel radiometer shows evidence of a thermally modulated signal outside of the eclipse season, and possibly some weaker evidence for variations associated with a voltage coupling. No other channel shows evidence for thermal or voltage modulated signals outside of the eclipse season. We place upper limits on these small residual systematic effects, after excluding 31 GHz data from the eclipse season, in §4.3.

### 4.4.2. Lock-In Amplifier Memory

Kogut et al. (1992) identified a small "memory" in the time-ordered data at an amplitude of about 3.2%, i.e. the datum in each half-second sample "remembers" 3.2% of the previous sample. This effect is due to the lock-in amplifiers that amplify and integrate the DMR signals. We correct for this effect by subtracting from each half-second datum 3.2% of the previous datum value. Power spectra of the time-ordered data are computed after the time-ordered data were corrected for our best estimates of the magnetic susceptibility, lunar emission, planetary emission, and Doppler and cosmic dipoles. A 15° Galactic cut is applied in this analysis. The power spectra are Fourier-transformed to produce the autocorrelation function from lag 0 to 512 points. The lock-in memory is clearly apparent at lag 0.5 s. Table 9 expresses the amplitude of the autocorrelation at 0.5 s lag relative to the autocorrelation at zero lag.

### 4.4.3. Attitude Errors

A coarse attitude pointing solution is calculated based on the data from the spacecraft's attitude control system. The uncertainty in the coarse attitude is less than $4'$ ($1\sigma$).

Fine attitude solutions for the *COBE* instruments are derived from the DIRBE. The DIRBE observations of stars are compared with known stellar positions, and fine attitude corrections are determined. An FFT analysis of DIRBE attitude residuals show no periodic effects near the orbital or spin frequency. The frequency spectrum is close to that of white noise. The residuals have an uncertainty of less than $2'$ ($1\sigma$). Greater than 99% of the attitude data used in our analysis are based on the fine attitude solutions. Overall, for 99% of the time the systematic pointing errors are much less than $3'$.

## 5. Results

The best fit dipole from the two year DMR data is $3.363 \pm 0.024$ mK towards Galactic coordinates $(\ell, b) = (264.4° \pm 0.2°, +48.1° \pm 0.4°)$ for $|b| > 15°$, in excellent agreement with the first year results of Kogut et al. (1993). The dipole is removed for all further analysis of the two year data, below.

Figures 1 and 2 show the microwave sky maps based on the first two years of DMR data. Fluctuations in the CMB are detected in a $10°$ patch with a signal-to-noise ratio that is greater than one in the two-year DMR data; this was not the case with the first year data. Still, the contribution of noise to the total signal is significant and plays an important role in statistical calculations. Neglecting systematic effects, each map pixel, $i$, has an observed temperature, $T_{obs,i}$, which is a result of a true CMB temperature, $T_{CMB,i}$, and a noise contribution, $T_{n,i}$,

$$T_{obs,i} = T_{CMB,i} + T_{n,i}. \tag{5}$$

For random Gaussian instrument noise the quadratic statistic

$$T_{obs,i}^2 = T_{CMB,i}^2 + T_{n,i}T_{CMB,i} + T_{n,i}^2 \tag{6}$$

has an expectation value of

$$\left\langle T_{obs,i}^2 \right\rangle = \left\langle T_{CMB,i}^2 \right\rangle + \left\langle T_{n,i}^2 \right\rangle \tag{7}$$

thus $T_{obs,i}^2$ is a biased estimator of $T_{CMB,i}^2$. This *noise bias* is significant and is not limited to the particular quadratic statistic noted above, but occurs in a variety of statistical analyses of the DMR maps (see, e.g. Smoot et al. 1994 and Hinshaw et al. 1994).

In general, a given cosmological model does not predict the exact CMB temperature that would be observed in our sky, rather it will predict a statistical distribution of anisotropy parameters, such as spherical harmonic amplitudes. In the context of such models, the true CMB temperature observed in our sky is only a single realization from a statistical distribution. Thus, in addition to experimental uncertainties, we must also assign a *cosmic variance* uncertainty to cosmological parameters derived from the DMR maps. It is important to recognize that cosmic variance exists independent of the quality of the experiment.

In the analyses discussed below we take into account experimental noise, systematic error upper limits, noise biases, and, where noted, cosmic variance.

## 5.1. RMS fluctuations

Table 23 shows the RMS thermodynamic temperature fluctuation values derived from the first year data, the second year data, and the first two years of data combined, as a function of Galactic cut angle. Table 24 shows the same information using the *combination* and *subtraction* techniques to reduce the Galactic signal, as described by Bennett et al. (1992b) and in §5.3 below. Results are given for both the unsmoothed 7° resolution data (for a precise characterization of the DMR window function see Wright et al. (1994a)) and for data smoothed with a 7° FWHM Gaussian kernel to a total effective angular resolution of 10°. The quadrupole is not removed and the kinematic quadrupole correction (discussed in §5.3) is not applied. We present separate results for the $(A + B)/2$ sum maps, $\sigma_{obs}$, and the $(A - B)/2$ difference maps, $\sigma_n$, the latter of which provides an estimate of the experimental noise. The best estimate of the RMS sky temperature fluctuations is

$$\sigma_{sky} = \sqrt{\sigma_{obs}^2 - \sigma_n^2}. \tag{8}$$

These results were derived using uniform weights for the sky pixels. In cases where instrument noise is the limiting factor in the RMS determination, weighting by the square of the number of observations, $n_{obs}^2$, is optimal, while in cases where cosmic variance is the limiting factor, uniform weighting is preferred. Although uniform weighting is not exactly optimal for determining the sky RMS for our two year data, it is nearly so and the results are more directly applicable to model comparisons. Error estimates are computed for the full two year data set using Monte Carlo simulations that include the effects of instrument noise and systematics, but not cosmic variance. For the relatively sensitive 53 GHz channels the RMS fluctuation amplitudes are $44 \pm 7$ $\mu$K and $30.5 \pm 2.7$ $\mu$K for the 7° and 10° resolution data, respectively. To account for cosmic variance an additional model-dependent $\sim 4$ $\mu$K ($Q_{rms-PS} = 17$ $\mu$K, $n = 1$) should be added in quadrature with the quoted uncertainties.

Table 1 of Smoot et al. (1992) presents the first year RMS temperature fluctuations smoothed to 10° resolution. The results for the first year reported here do not agree precisely with those in Smoot et al. (1992) because: (1) software analysis modifications were implemented, (2) results in Table 1 of Smoot et al. (1992) were calculated with pixel weights equal to the number of observations per pixel, and (3) some of the 31 GHz entries are in error in Smoot et al.'s Table 1. The overall results are consistent, however.

## 5.2. Two-point correlations

The two-point 53 GHz $(A + B)/2 \times$ 90 GHz $(A + B)/2$ cross correlation function is shown in Figure 3. It has the same general appearance as the first year function, but the uncertainty in the application of these data to cosmological models is now almost entirely dominated by cosmic variance, particularly in the quadrupole moment, rather than by instrument noise or systematic errors. The zero lag amplitude of the cross-correlation function (including the quadrupole) is $C(0)^{1/2} = 36 \pm 5$ $\mu$K (68% CL) for the 7° resolution data, not including cosmic variance in the uncertainty. Assuming a power law model of initial Gaussian density fluctuations, $P(k) \propto k^n$, we determine the most likely quadrupole normalized amplitude, $Q_{rms-PS}$, and spectral index, $n$, by evaluating the Gaussian approximation to the likelihood function, $L(Q_{rms-PS}, n)$, as defined in equation 1 of Seljak & Bertschinger (1993). We estimate mean correlation functions and covariance matrices for a range of $Q_{rms-PS}$ and $n$ values by means of Monte Carlo simulations that account for all important aspects of our data processing including monopole and dipole (and quadrupole) subtraction on the cut sky. The derived covariance matrices are inverted using singular value decomposition which permits an unambiguous identification of the zero modes that arise due to multipole subtraction. Figure 4 shows the resulting likelihood contours as a function of $Q_{rms-PS}$ and $n$ for the analyses with and without the quadrupole anisotropy. The contours correspond to 68%, 95%, and 99.7% confidence regions, as obtained by direct integration of the likelihood function. The most likely values for $Q_{rms-PS}$ and $n$ are $12.4^{+5.2}_{-3.3}$ $\mu$K (68% CL) and $1.59^{+0.49}_{-0.55}$ (68% CL), including the quadrupole, where the quoted uncertainties encompass the 68% confidence region in two dimensions and include cosmic variance. With $n$ fixed to unity the most likely quadrupole-normalized amplitude is $17.4 \pm 1.5$ $\mu$K (68% CL). Excluding the quadrupole anisotropy, the most likely values for $Q_{rms-PS}$ and $n$ are $16.0^{+7.5}_{-5.2}$ $\mu$K (68% CL) and $1.21^{+0.60}_{-0.55}$ (68% CL), and with $n$ fixed to unity the most likely quadrupole-normalized amplitude is $18.2 \pm 1.6$ $\mu$K (68% CL).

Monte Caro simulations were performed in order to identify possible biases in our statistical technique which may arise, e.g., from the Gaussian approxmation we employ, or from numerical uncertainty in the Monte Carlo determination of the mean correlation functions and covariance matrices. We have generated a sample of 3000 cross-correlation functions computed from sky maps with a quadrupole-normalized amplitude of 18 $\mu$K and spectral index $n = 1$ (including appropriate instrument noise). For each function in the sample we evaluated the Gaussian likelihood to determine the most likely values of $Q_{rms-PS}$ and $n$. The resulting ensemble of most-likely $n$ values had a mean of 1.1 and a standard deviation of 0.6. Thus we conclude there is a bias of $\sim 0.1$ in $n$, however, since this bias is much smaller than the spread in $n$ induced by cosmic variance and noise, we have NOT corrected the most-likely values reported above for this effect. To complete this test, we have selected from our sample, the subset of maps in which the actual quadrupole moment was close to the low value observed in our sky (between 3 and 9 $\mu$K, see §5.3). The resulting subset of most-likely $n$ values had a mean of $\sim 1.3$. Thus the most likely bias-corrected estimate of $n$ is between 1.1 and 1.3.

Clearly the amplitude and spectral index are not separately well constrained by the two year COBE data. A high quadrupole value is better fit with a small value of $n$, and vice versa, thus there is a correlated ridge in the likelihood of $n$ and $Q_{rms-PS}$. The ridge of maximum likelihood depicted in Figure 4 is well described by the relation

$$Q_{rms-PS} = 17.4\ e^{0.58(1-n)}\ \mu K. \tag{9}$$

In fact, there exists a pivot point in the power sectrum where the multipole amplitude is independent of $n$. For the above determination of the power spectrum parameters we find that the pivot point occurs at spherical harmonic order $\ell = 7$ ($a_7 \cong 9.5\ \mu$K), while Górski et al. (1994) deduce a pivot at $\ell = 9$. This may suggest that the two-point correlation function probes less deeply into the power spectrum than the technique employed by Górski et al. (1994).

Scaramella & Vittorio (1993) perform a $\chi^2$ minimization with Monte Carlo simulations of the effects of cosmic variance on the first year DMR data and deduce $Q_{rms-PS} = (14.5 \pm 1.7)(1 \pm 0.06)\ \mu$K, but the actual DMR sky sampling and data reduction technique were not taken into account. Seljak & Bertschinger analyze the first year DMR data using a maximum likelihood technique and conclude $Q_{rms-PS} = (15.7 \pm 2.6)\exp[0.46(1 - n)]\ \mu$K. Smoot et al. (1994) analyze the first year DMR data using an rms fluctuation amplitude versus smoothing angle statistic to arrive at $Q_{rms-PS} = (13.2 \pm 2.5)$ and $n = 1.7^{+0.3}_{-0.6}$.

Wright et al. (1994b) solve for the angular power spectrum of the DMR two year data by modifying and applying the technique described by Peebles (1973) and Hauser & Peebles (1973) for data on a cut sphere. For $\ell$ from 3 to 19, Wright et al. conclude that the data are well-described by the power spectrum $P(k) \propto k^n$ where $n = 1.46^{+0.39}_{-0.44}$, with $n = 1$ only $1\sigma$ from the best fit value. Fixing $n = 1$ results in fluctuation amplitudes of $Q_{rms-PS} = 19.6 \pm 2.0, 19.3 \pm 1.3$, and $16.0 \pm 2.1\ \mu$K for the 53, 53+90, and Galaxy removed data, respectively. If the $\ell$-range is extended to 30, then Wright et al. derive $n = 1.25^{+0.4}_{-0.45}$.

Górski et al. (1994) construct functions that are orthogonal on the cut sphere and form an exact likelihood directly in terms of these functions. They analyze the 53 and 90 GHz data concurrently and arrive at maximum likelihood values of $Q_{rms-PS} = 17.0 \ \mu$K and $n = 1.22$. Excluding the quadrupole Górski et al. (1994) derive maximum likelihood values of $Q_{rms-PS} = 20.0 \ \mu$K and $n = 1.02$.

There is great interest in using the *COBE* DMR data to discriminate between Gaussian and non-Gaussian cosmological models. Unfortunately, given the limitations of the DMR noise level, statistical noise bias, and cosmic variance, it is difficult to constrain non-Gaussian models. Hinshaw et al. (1994) and Smoot et al. (1994) discuss this in the context of the first year DMR data. They find that the first year DMR data are consistent with Gaussian CMB fluctuation statistics, but do not test (and thus do not rule out) particular non-Gaussian cosmological models.

### 5.3. Galaxy removal and the quadrupole

The dipole and quadrupole moments are the most susceptible spherical harmonic modes to contamination by experimental systematic errors including local Galactic emission. Bennett et al. (1992b) showed that the Galactic quadrupole is significant and must be accurately modeled before a cosmic quadrupole can be estimated. Figure 4 of Bennett et al. (1992b) shows that the Galactic signal can be largely ignored for smaller angular scale fluctuations after a Galactic cut is applied.

Bennett et al. (1992b) defined three techniques for separating cosmic and Galactic emission: a *subtraction technique*, where externally measured Galactic signals are extrapolated in frequency and subtracted from the DMR maps; a *fit technique*, which fits models of the Galactic and cosmic emission to the DMR data after a portion of the Galactic signal has been modeled and removed; and a *combination technique* that relies only on linear combinations of the DMR data with no use of external data. (The numbers 1.523 and 1.143 in the text of §5.1 of Bennett et al. (1992b) should read 1.636 and 1.401, respectively, although this does not change any of their results.) Since the subtraction and the fit techniques give rise to nearly identical CMB maps, we only include results derived from the subtraction and combination technique CMB maps in this paper. The subtraction technique CMB map may be expressed as a linear combination of data from the six DMR channels after subtracting external synchrotron and dust emission models:

$$T_{Sub} = -0.341 \times \frac{1}{2}(T'_{31A} \pm T'_{31B}) + 0.817 \times \frac{1}{2}(T'_{53A} \pm T'_{53B}) + 0.701 \times \frac{1}{2}(T'_{90A} \pm T'_{90B}) \quad (10)$$

where $T'$ is the DMR map temperature after subtraction of the synchrotron and dust emission models. The combination technique CMB map is a linear combination of un-subtracted channel maps, with weights that depend on the amplitudes of the signals in the Galactic plane. Following Bennett et al. (1992b), for the two-year data the average signals in the outer galactic plane ($\sin|b| < 0.1$ and $|\ell| > 30°$) are $T^G = 1.259$, $T^G = 0.393$, and $T^G = 0.262$, so we have

$$T_{Comb} = -0.479 \times \frac{1}{2}(T_{31A} \pm T_{31B}) + 1.393 \times \frac{1}{2}(T_{53A} \pm T_{53B}) + 0.207 \times \frac{1}{2}(T_{90A} \pm T_{90B}). \quad (11)$$

Note that the combination map is somewhat noisier than the subtraction map.

We form a map of free-free emission (in antenna temperature at 53 GHz) using the following linear combination of the synchrotron- and dust-subtracted 31 and 53 GHz maps

$$T^{ff} = +0.499 \times \frac{1}{2}(T'_{31A} \pm T'_{31B}) - 0.522 \times \frac{1}{2}(T'_{53A} \pm T'_{53B}). \quad (12)$$

The amplitude of the free-free continuum emission derived from the two-year data is $T_A(\mu K) = 10 \pm 4 \csc|b|$ at 53 GHz for $|b| > 15°$. The peak Galactic signal amplitudes in the 31, 53, and 90 GHz maps are $5.74 \pm 0.25$ mK towards $(\ell, b) = (334.7°, -1.2°)$, $1.76 \pm 0.08$ mK towards $(\ell, b) = (78.2°, 1.3°)$, and $1.16 \pm 0.17$ mK towards $(\ell, b) = (1.3°, 1.3°)$, respectively.

After modeled Galactic emission is removed from the DMR maps, it is still necessary to apply a Galactic plane cut before cosmological analysis of the data since even the residual Galactic signal can be significant in the plane. For the quadrupole analysis that follows we reject data where $|b| < 10°$. We define the five quadrupole components $Q_i$ by the expansion

$$\begin{aligned} Q(l, b) = \quad & Q_1(3\sin^2 b - 1)/2 + Q_2 \sin 2b \cos l + Q_3 \sin 2b \sin l + \\ & Q_4 \cos^2 b \cos 2l + Q_5 \cos^2 b \sin 2l \end{aligned} \quad (13)$$

where $l$ and $b$ are Galactic longitude and latitude, respectively. The RMS quadrupole amplitude is given by

$$Q_{rms}^2 = \frac{4}{15}\left[\frac{3}{4}Q_1^2 + Q_2^2 + Q_3^2 + Q_4^2 + Q_5^2\right]. \quad (14)$$

The second order Doppler term constitutes a kinematic quadrupole with an amplitude of $Q_{rms} = 1.2$ $\mu K$ and components $(Q_1, Q_2, Q_3, Q_4, Q_5) = (0.9, -0.2, -2.0, -0.9, 0.2)$ $\mu K$ (Bennett et al. 1992b). When the Galactic cut is applied the spherical harmonics are no longer orthogonal so uncertainties in the quadrupole arise from the aliasing of higher order spherical harmonic power onto the quadrupole. We compute these uncertainties by means of Monte Carlo simulations. For a given value of $Q_{rms-PS}$ and $n$ we simulate signals filtered through the DMR beam, record the true components that went into each given sky, denoted $q_i$, then fit a spherical harmonic expansion to $\ell = 2$ on the cut sky. This yields a set of recovered quadrupole components on the cut sky, denoted $Q_i$. The RMS uncertainty in the $q_i$'s due to aliasing is then defined to be:

$$\sigma_i = \sqrt{\langle (Q_i - q_i)^2 \rangle} \tag{15}$$

For a model with $Q_{rms-PS} = 17$ $\mu$K and $n = 1$, a $10°$ Galaxy cut, and uniform pixel weights, we find RMS uncertainties of 3.4, 0.4, 0.4, 3.6, and 3.5 $\mu$K (thermodynamic) for $Q_1$ through $Q_5$, respectively.

We form combined quadrupole uncertainties by taking a quadrature sum of the systematic, noise, and alias uncertainties for each CMB map considered. Simulations show that the choice of a $10°$ Galactic cut angle, uniform pixel weights, and a truncation of the spherical harmonic fit to order $\ell = 2$ combine to minimize the total uncertainty due to Galactic contamination, aliasing, and instrument noise. The best-fit quadrupole components and uncertainties are given in Tables 25 and 26. (The uncertainties quoted for the difference maps are smaller than the sum maps because they do not include the effects of aliasing.) In these Tables we define

$$\chi^2 = \sum_{i=1}^{5} Q_i^2 / \delta Q_i^2 \tag{16}$$

where $\delta Q_i$ are the uncertainties on the quadrupole components $Q_i$. Note that $\chi^2$ of the first year sum map indicates a significant (98% confidence) detection of signal relative to the component uncertainties. In contrast, $\chi^2$ limited to the second year of data is easily consistent with a zero quadrupole. Taken as a whole, the two year DMR data indicate a marginally significant (90% confidence) detection of a quadrupole signal.

An obvious question is whether or not the first and second year data are mutually consistent within their errors. We address this by examining the $\chi^2$ of the difference between the first and second year maps,

$$\chi^2 = \sum_{i=1}^{5} \frac{(Q1_i - Q2_i)^2}{\delta Q1_i^2 + \delta Q2_i^2} \tag{17}$$

where the $\delta Q$ values are based on the difference maps and do not include the aliasing errors. For the subtraction technique $\chi^2 = 7.1$ and for the noisier combination technique $\chi^2 = 5.2$ for five degrees of freedom, so we conclude that the first and second year of data are reasonably consistent with one another.

In the limit of low signal-to-noise, which applies for the quadrupole analysis because of the need to use the relatively noisy 31 GHz data to remove the Galactic signal, the values of $Q_{rms}$ given in Tables 25 and 26 will generally overestimate the quadrupole power because of noise biasing and aliasing. We use Monte Carlo techniques to define a likelihood function for $Q_{rms}$. For a given set of observed quadrupole components $Q_i$ and uncertainties $\delta Q_i$ we define a set of true $Q_{rms}$ values along the ray towards the observed $Q_i$ values. For each true $Q_{rms}$ from this set we generate 50,000 simulated quadrupoles by adding noise distributed according to the observed $\delta Q_i$ values. We tabulate the number of realizations where the observed $Q_{rms}$ is within 0.5 $\mu$K of the simulated (noisy) $Q_{rms}$ and the $\chi^2$ of the simulation is at least as large as the $\chi^2$ of the observation (i.e. we require the simulated quadrupoles to be at least as significant as the observed quadrupole). We define the likelihood to be proportional to the fraction of realizations that satisfy these conditions for each value of the true $Q_{rms}$. The smoothed likelihood function for $Q_{rms}$ for the first year of DMR data is given in Figure 5a. We note that there is greater than 98% likelihood that a non-zero quadrupole is detected in this data. The likelihood function for the second year of data is shown in Figure 5b, where the detection of a non-zero quadrupole is only marginally likely. While the most likely quadrupole values are smaller in the second year of data than in the first, the difference is not statistically exceptional. The likelihood function for the full two year data set is shown in Figure 5c. We conclude that there is a quadrupole whose amplitude is $Q_{rms} \approx 6 \pm 3$ $\mu$K (68% CL), based on the first two years of DMR data. We note that 35% of the first year likelihood function and 45% of the second year likelihood function lies within this confidence interval. We also note that our initial first year quadrupole estimate of $13 \pm 4$ $\mu$K did not include corrections for noise biasing and aliasing effects. Both of these cause the mean $Q_{rms}$ to be overestimated, but not by a large amount, as seen in Figure 5a.

The first and second year data are reasonably consistent with each other, and there is likely to be a non-zero quadrupole consistent with all of the DMR data at a level of $Q_{rms} = 6 \pm 3$ $\mu$K (68% CL). There is no doubt that $Q_{rms}$ has a lower value than the quadrupole-normalization of the entire power spectrum, $Q_{rms-PS}$. Whether this is due to cosmic variance, Galactic model error, or reflects the cosmology of the universe remains to be determined. The probability of measuring a quadrupole of amplitude $3 < Q_{rms}$ ($\mu$K) $< 9$ from a power spectrum normalized to $Q_{rms-PS} = 17$ $\mu$K is 10%.

We emphasize that the *COBE* quadrupole uncertainty results in part from the need to use the relatively noisy 31 GHz data to remove a portion of the Galactic signal. This is true for all three of the Galactic removal techniques that have been used. The analysis of all four years of DMR data will not improve this uncertainty significantly because only one of the 31 GHz channels contains useful data for the second two years. Alternative techniques to measure the free-free emission would be invaluable in reducing the uncertainty in the quadrupole determination made by COBE. The *Wisconsin Hα Mapper (WHAM)*, now under construction, would have the capability to provide an accurate estimate of the free-free sky emission with 0.8° resolution, but a map of the entire sky would be needed. Full sky absolute radio continuum surveys at frequencies greater than 10 to 15 GHz would complement this effort and facilitate a cleaner separation of the free-free emission from the Galactic synchrotron emission.

## 6. Summary

1. The methods and execution of the processing of the first two years of DMR data are discussed, and the data calibration and systematic error upper limits are presented.

2. The first year and second year results are consistent with each other.

3. The RMS CMB temperature fluctuations at 7° and 10° angular resolution are $44 \pm 7 \mu K$ and $30.5 \pm 2.7 \mu K$, respectively, for the 53 GHz data. The cross correlation of the 53 GHz data with the 90 GHz data at zero lag gives another estimate of the RMS fluctuations at 7° angular resolution. We find $C(0)^{1/2} = 36 \pm 5 \mu K$ (68% CL), with the quadrupole included. The uncertainty does not include cosmic variance.

4. We present the two point cross correlation function of the 53 and 90 GHz DMR data. The best estimate of the power spectrum amplitude, with the quadrupole included, is $Q_{rms-PS} = 12.4^{+5.2}_{-3.3} \mu K$ (68% CL) with a spectral index of $n = 1.59^{+0.49}_{-0.55}$ (68% CL). For $n$ fixed to unity we find $Q_{rms-PS} = 17.4 \pm 1.5 \mu K$ (68% CL). With the quadrupole excluded we get $Q_{rms-PS} = 16.0^{+7.5}_{-5.2} \mu K$ (68% CL) and $n = 1.21^{+0.60}_{-0.55}$ (68% CL). For $n$ fixed to unity we find $Q_{rms-PS} = 18.2 \pm 1.6 \mu K$ (68% CL). Monte Carlo simulations indicate that these derived estimates of $n$ may be biased by $\sim +0.3$ (with the observed low value of the quadrupole included in the analysis) and $\sim +0.1$ (with the quadrupole excluded). Thus the most likely bias-corrected estimate of $n$ is between 1.1 and 1.3. These results are consistent with those derived by Wright et al. (1994b) and Górski et al. (1994).

5. The best dipole determination from the two-year DMR data is $3.363 \pm 0.024$ mK towards Galactic coordinates $(\ell, b) = (264.4° \pm 0.2°, +48.1° \pm 0.4°)$, in excellent agreement with the first year results of Kogut et al. (1993) and Fixsen et al. (1994).

6. The quadrupole in the second year of data is smaller than our best estimate from the first year of data alone (although the likelihood functions are not inconsistent). The addition of the second year of data has reduced our best estimate of the RMS quadrupole to $Q_{rms} = 6 \pm 3 \mu K$ (68% CL).

We gratefully acknowledge the many people who made this paper possible: the NASA

Table 1: DMR Sky Map Coverage and Antenna Temperature Sensitivity Limits

| DMR Channel | 31A | 31B | 53A | 53B | 90A | 90B |
|---|---|---|---|---|---|---|
| flight RMS (mK in 0.5 s) | 58.6 | 60.4 | 23.2 | 27.1 | 39.7 | 30.2 |
| Max obs per pixel | 71145 | 62839 | 81271 | 81400 | 81490 | 81669 |
| Noise RMS per pixel (mK) | 0.220 | 0.241 | 0.081 | 0.095 | 0.139 | 0.106 |
| Mean obs per pixel | 29728 | 24484 | 35672 | 35681 | 35655 | 35643 |
| Noise RMS per pixel (mK) | 0.340 | 0.386 | 0.123 | 0.143 | 0.210 | 0.160 |
| Min obs per pixel | 11506 | 7988 | 21526 | 21471 | 21579 | 21557 |
| Noise RMS per pixel (mK) | 0.546 | 0.676 | 0.158 | 0.185 | 0.270 | 0.206 |

Table 2: Comparisons of the Noise Source Gain Determination[a]

| Chan | Ground (%) | Flight (%) | Rel. A/B (%) | Linear Drift (% yr$^{-1}$) | Orbit Drift ($\frac{\Delta G}{G} \times 10^5$) | Spin Drift ($\frac{\Delta G}{G} \times 10^7$) |
|---|---|---|---|---|---|---|
| 31A | $0.0 \pm 2.5$ | $+2.1 \pm 3.5$ | $0.41 \pm 0.06$ | $-0.01 \pm 0.07$ | 10.0 | 2.6 |
| 31B | $0.0 \pm 2.3$ | $+1.7 \pm 3.5$ | | $+0.10 \pm 0.09$ | 18.4 | 6.3 |
| 53A | $0.0 \pm 0.7$ | $-2.2 \pm 1.8$ | $0.21 \pm 0.02$ | $+0.01 \pm 0.03$ | 4.4 | 2.9 |
| 53B | $0.0 \pm 0.7$ | $-1.4 \pm 2.0$ | | $-0.06 \pm 0.03$ | 5.3 | 2.3 |
| 90A | $0.0 \pm 2.0$ | $+6.3 \pm 3.6$ | $0.16 \pm 0.04$ | $+0.15 \pm 0.05$ | 9.1 | 3.7 |
| 90B | $0.0 \pm 1.2$ | $-0.9 \pm 2.5$ | | $+0.18 \pm 0.04$ | 9.6 | 2.9 |

[a]The mean drifts have *not* been applied as corrections. All uncertainties are 68% confidence except for the spin and orbit drifts, which are 95% confidence upper limits.

Table 3: Absolute Gain Comparisons Relative to the Noise-Source Gain Determination

| Channel | Ground | Doppler | Moon |
|---------|--------|---------|------|
| 31A | $1.000 \pm 0.025$ | $1.090 \pm 0.045$ | $1.021 \pm 0.035$ |
| 31B | $1.000 \pm 0.023$ | $1.067 \pm 0.053$ | $1.017 \pm 0.035$ |
| 53A | $1.000 \pm 0.007$ | $0.978 \pm 0.018$ | $1.021 \pm 0.054$ |
| 53B | $1.000 \pm 0.007$ | $0.986 \pm 0.020$ | $1.019 \pm 0.054$ |
| 90A | $1.000 \pm 0.020$ | $1.063 \pm 0.036$ | $1.013 \pm 0.063$ |
| 90B | $1.000 \pm 0.013$ | $0.991 \pm 0.025$ | $1.012 \pm 0.063$ |

Table 4: Comparisons of A/B Noise Source Gain Determination

| Frequency | Ground | Dipole | Moon |
|-----------|--------|--------|------|
| 31 | $1.000 \pm 0.026$ | $1.021 \pm 0.070$ | $1.0041 \pm 0.0006$ |
| 53 | $1.000 \pm 0.003$ | $0.992 \pm 0.027$ | $1.0021 \pm 0.0002$ |
| 90 | $1.000 \pm 0.010$ | $1.073 \pm 0.044$ | $1.0016 \pm 0.0004$ |

Table 5: Calibration Drifts

| Channel | Noise Source Ratios ($\%~\mathrm{yr}^{-1}$) | Gain Drift from Dipole ($\%~\mathrm{yr}^{-1}$) | Gain Drift from Moon ($\%~\mathrm{yr}^{-1}$) |
|---------|---------------------------------------------|------------------------------------------------|----------------------------------------------|
| 31A | $+0.111 \pm 0.009$ | $+0.34 \pm 0.74$ | $-0.01 \pm 0.07$ |
| 31B | $-0.314 \pm 0.009$ | $-0.05 \pm 1.13$ | $+0.10 \pm 0.09$ |
| 53A | $+0.056 \pm 0.002$ | $-0.02 \pm 0.38$ | $+0.011 \pm 0.027$ |
| 53B | $+0.160 \pm 0.002$ | $-0.23 \pm 0.42$ | $-0.057 \pm 0.031$ |
| 90A | $+0.181 \pm 0.006$ | $+0.09 \pm 0.74$ | $+0.15 \pm 0.05$ |
| 90B | $+0.287 \pm 0.005$ | $+0.49 \pm 0.53$ | $+0.18 \pm 0.04$ |

Table 6: Magnetic Susceptibility: Antenna Temperature per Applied Magnetic Field Strength

| Channel | $\beta_X$ (mK Gauss$^{-1}$) | $\beta_R$ (mK Gauss$^{-1}$) | $\beta_T$ (mK Gauss$^{-1}$) |
|---------|------------------|------------------|------------------|
| 31A | $-0.3624 \pm 0.0209$ | $+0.2614 \pm 0.0525$ | $-0.1759 \pm 0.0884$ |
| 31B | $+0.1472 \pm 0.0818$ | $+0.2496 \pm 0.0628$ | $+0.0644 \pm 0.1011$ |
| 53A | $-1.5122 \pm 0.0079$ | $-0.1030 \pm 0.0262$ | $-0.9259 \pm 0.0312$ |
| 53B | $+0.0870 \pm 0.0088$ | $-0.3370 \pm 0.0399$ | $-0.2004 \pm 0.0345$ |
| 90A | $-0.1241 \pm 0.0142$ | $-1.1021 \pm 0.0332$ | $-0.3243 \pm 0.0558$ |
| 90B | $+0.0044 \pm 0.0100$ | $+0.1236 \pm 0.0234$ | $-0.1765 \pm 0.0392$ |

Table 7: Diffraction Model Antenna Temperatures Fitted To Earth-Binned Maps

| Channel | Earth Above Shield ($\mu$K) | Earth Below Shield ($\mu$K) |
|---------|------------------|------------------|
| 31A | $4 \pm 13$ | $31 \pm 31$ |
| 31B | $109 \pm 53$ | $29 \pm 32$ |
| 53A | $51 \pm 26$ | $5 \pm 12$ |
| 53B | $96 \pm 28$ | $17 \pm 14$ |
| 90A | $139 \pm 35$ | $7 \pm 18$ |
| 90B | $164 \pm 35$ | $7 \pm 16$ |

Table 8: 95% Confidence Upper Limits to the Antenna Temperature of Earth Signal Contribution to the DMR Data

| Channel | Current 2-Year Limits ($\mu$K) | Previous 1-Year Limits ($\mu$K) |
|---------|--------------------------------|---------------------------------|
| 31A | 62 | 130 |
| 31B | 62 | 135 |
| 53A | 24 | 57 |
| 53B | 28 | 47 |
| 90A | 36 | 102 |
| 90B | 32 | 88 |

Table 9: Amplitude of the Lock-in Memory Systematic Error Before Correction[a]

| Channel | Lock-in Memory Amplitude (% of signal) |
|---------|----------------------------------------|
| 31A | $3.192 \pm 0.004$ |
| 31B | $3.149 \pm 0.005$ |
| 53A | $3.204 \pm 0.004$ |
| 53B | $3.181 \pm 0.003$ |
| 90A | $3.119 \pm 0.004$ |
| 90B | $3.127 \pm 0.003$ |

[a]Uncertainties are statistical only, with no contribution from systematics.

Table 10: 95% Confidence Antenna Temperature Upper Limits from Orbit- and Spin-Modulated Effects

| Channel | Orbit-Modulated Effects | | | Spin-Modulated Effects | |
|---|---|---|---|---|---|
| | Eclipse | FFT | Orbit Binning | FFT | Spin Binning |
| | ($\mu$K) | ($\mu$K) | ($\mu$K) | ($\mu$K) | ($\mu$K) |
| 31A | 101 | 392 | 99 | 502 | 22 |
| 31B | 31 | 309 | 146 | 516 | 47 |
| 53A | 18 | 91 | 28 | 185 | 10 |
| 53B | 49 | 98 | 19 | 205 | 15 |
| 90A | 37 | 150 | 38 | 286 | 19 |
| 90B | 9 | 109 | 66 | 216 | 10 |

Table 11: Quadrature Sum of Systematic Effects for Channel 31A

| Effect | P-P[a] | RMS[b] | $\Delta T_1$ | $\Delta T_2$ | $\Delta T_3$ | $\Delta T_4$ | $\Delta T_5$ | $\Delta T_6$ | $\Delta T_7$ | $\Delta T_8$ |
|---|---|---|---|---|---|---|---|---|---|---|
| | ($\mu$K) | ($\mu$K) | ($\mu$K) | ($\mu$K) | ($\mu$K) | ($\mu$K) | ($\mu$K) | ($\mu$K) | ($\mu$K) | ($\mu$K) |
| | | | | | | | | | | |
| Channel 31A Before Correction[c] | | | | | | | | | | |
| $\beta_X$ | 4.92 | 0.55 | 0.15 | 0.15 | 0.04 | 0.08 | 0.07 | 0.03 | 0.02 | 0.02 |
| $\beta_R$ | 33.59 | 6.61 | 1.27 | 5.67 | 0.65 | 2.93 | 0.35 | 0.56 | 0.13 | 0.70 |
| $\beta_T$ | 17.87 | 3.01 | 18.74 | 1.57 | 2.07 | 0.57 | 0.90 | 0.25 | 0.18 | 0.14 |
| Earth | 4.14 | 0.47 | 0.41 | 0.28 | 0.14 | 0.20 | 0.14 | 0.12 | 0.02 | 0.09 |
| Moon | 10.12 | 1.69 | 0.05 | 1.25 | 0.05 | 0.43 | 0.03 | 0.26 | 0.03 | 0.52 |
| Doppler | 83.24 | 12.74 | 67.74 | 6.14 | 8.83 | 1.88 | 4.64 | 1.43 | 0.79 | 0.93 |
| Spin | 6.00 | 0.75 | 0.23 | 0.37 | 0.19 | 0.30 | 0.17 | 0.25 | 0.03 | 0.15 |
| Other | 99.35 | 11.92 | 0.62 | 0.78 | 0.40 | 0.62 | 0.41 | 0.57 | 0.49 | 0.64 |
| TOTAL[d] | 135.75 | 19.00 | 70.30 | 8.65 | 9.10 | 3.62 | 4.76 | 1.70 | 0.95 | 1.44 |
| | | | | | | | | | | |
| Channel 31A After Correction[e] | | | | | | | | | | |
| $\beta_X$ | 0.81 | 0.09 | 0.02 | 0.02 | 0.01 | 0.01 | 0.01 | 0.00 | 0.00 | 0.00 |
| $\beta_R$ | 14.04 | 2.76 | 0.53 | 2.37 | 0.27 | 1.22 | 0.14 | 0.23 | 0.05 | 0.29 |
| $\beta_T$ | 18.09 | 3.05 | 18.96 | 1.59 | 2.09 | 0.58 | 0.91 | 0.25 | 0.18 | 0.14 |
| Earth | 7.34 | 0.83 | 0.72 | 0.50 | 0.24 | 0.35 | 0.24 | 0.22 | 0.04 | 0.15 |
| Moon | 1.09 | 0.18 | 0.01 | 0.14 | 0.01 | 0.05 | 0.00 | 0.03 | 0.00 | 0.06 |
| Doppler | 4.19 | 0.64 | 3.39 | 0.31 | 0.44 | 0.10 | 0.23 | 0.07 | 0.04 | 0.05 |
| Spin | 13.32 | 1.67 | 0.51 | 0.81 | 0.42 | 0.67 | 0.38 | 0.56 | 0.06 | 0.33 |
| Other | 8.72 | 0.61 | 0.25 | 0.12 | 0.11 | 0.11 | 0.10 | 0.10 | 0.08 | 0.11 |
| TOTAL[d] | 29.17 | 4.61 | 19.29 | 3.03 | 2.21 | 1.56 | 1.06 | 0.70 | 0.22 | 0.50 |

[a]The peak-to-peak amplitude in the map (after the best-fit dipole has been removed).

[b]The pixel-to-pixel standard deviation (after the best-fit dipole has been removed).

[c]Uncertainties refer to antenna temperature of the individual channel or $(A + B)/2$ "sum" maps.

[d]The Total refers to the sum, in quadrature, of the individual effects.

[e]After correction $\Delta T$ terms are antenna temperature 95% confidence level upper limits.

Table 12: Quadrature Sum of Systematic Effects for Channel 31B

| Effect | P-P[a] | RMS[b] | $\Delta T_1$ | $\Delta T_2$ | $\Delta T_3$ | $\Delta T_4$ | $\Delta T_5$ | $\Delta T_6$ | $\Delta T_7$ | $\Delta T_8$ |
|---|---|---|---|---|---|---|---|---|---|---|
| | ($\mu$K) | ($\mu$K) | ($\mu$K) | ($\mu$K) | ($\mu$K) | ($\mu$K) | ($\mu$K) | ($\mu$K) | ($\mu$K) | ($\mu$K) |
| | | | | | | | | | | |
| Channel 31B Before Correction[c] | | | | | | | | | | |
| $\beta_X$ | 1.99 | 0.22 | 0.03 | 0.08 | 0.02 | 0.04 | 0.02 | 0.01 | 0.01 | 0.01 |
| $\beta_R$ | 44.19 | 8.54 | 1.70 | 7.44 | 1.27 | 3.29 | 0.65 | 0.89 | 0.36 | 0.67 |
| $\beta_T$ | 6.89 | 1.26 | 6.86 | 0.53 | 0.91 | 0.28 | 0.42 | 0.10 | 0.08 | 0.07 |
| Earth | 4.85 | 0.55 | 0.49 | 0.32 | 0.13 | 0.26 | 0.15 | 0.15 | 0.03 | 0.10 |
| Moon | 11.11 | 1.72 | 0.10 | 1.26 | 0.09 | 0.43 | 0.05 | 0.27 | 0.05 | 0.53 |
| Doppler | 114.64 | 16.76 | 69.67 | 7.67 | 12.26 | 3.61 | 5.60 | 1.89 | 0.98 | 1.29 |
| Spin | 6.00 | 0.75 | 0.23 | 0.37 | 0.19 | 0.30 | 0.17 | 0.25 | 0.03 | 0.15 |
| Other | 135.42 | 13.84 | 0.68 | 0.82 | 0.51 | 0.73 | 0.49 | 0.72 | 0.67 | 0.81 |
| TOTAL[d] | 183.49 | 23.47 | 70.03 | 10.82 | 12.37 | 4.98 | 5.68 | 2.25 | 1.25 | 1.75 |
| | | | | | | | | | | |
| Channel 31B After Correction[e] | | | | | | | | | | |
| $\beta_X$ | 2.22 | 0.24 | 0.03 | 0.09 | 0.02 | 0.04 | 0.02 | 0.02 | 0.01 | 0.01 |
| $\beta_R$ | 22.80 | 4.41 | 0.88 | 3.84 | 0.65 | 1.70 | 0.34 | 0.46 | 0.19 | 0.34 |
| $\beta_T$ | 21.66 | 3.96 | 21.56 | 1.66 | 2.87 | 0.89 | 1.31 | 0.33 | 0.26 | 0.21 |
| Earth | 9.01 | 1.02 | 0.92 | 0.59 | 0.23 | 0.48 | 0.28 | 0.28 | 0.06 | 0.19 |
| Moon | 1.18 | 0.18 | 0.01 | 0.13 | 0.01 | 0.05 | 0.00 | 0.03 | 0.00 | 0.06 |
| Doppler | 5.31 | 0.78 | 3.20 | 0.36 | 0.56 | 0.17 | 0.26 | 0.09 | 0.05 | 0.06 |
| Spin | 28.60 | 3.58 | 1.10 | 1.75 | 0.90 | 1.44 | 0.82 | 1.20 | 0.12 | 0.71 |
| Other | 12.44 | 0.97 | 0.31 | 0.14 | 0.14 | 0.13 | 0.12 | 0.13 | 0.12 | 0.14 |
| TOTAL[d] | 45.58 | 7.11 | 21.86 | 4.59 | 3.14 | 2.45 | 1.63 | 1.37 | 0.37 | 0.85 |

[a]The peak-to-peak amplitude in the map (after the best-fit dipole has been removed).

[b]The pixel-to-pixel standard deviation (after the best-fit dipole has been removed).

[c]Uncertainties refer to antenna temperature of the individual channel or $(A + B)/2$ "sum" maps. All $\Delta T$ terms are 95% confidence level upper limits.

[d]The Total refers to the sum, in quadrature, of the individual effects.

[e]After correction $\Delta T$ terms are antenna temperature 95% confidence level upper limits.

Table 13: Quadrature Sum of Systematic Effects for Channel 53A

| Effect | P-P[a] | RMS[b] | $\Delta T_1$ | $\Delta T_2$ | $\Delta T_3$ | $\Delta T_4$ | $\Delta T_5$ | $\Delta T_6$ | $\Delta T_7$ | $\Delta T_8$ |
|---|---|---|---|---|---|---|---|---|---|---|
| | ($\mu$K) | ($\mu$K) | ($\mu$K) | ($\mu$K) | ($\mu$K) | ($\mu$K) | ($\mu$K) | ($\mu$K) | ($\mu$K) | ($\mu$K) |
| | | | | | | | | | | |
| | | | Channel 53A Before Correction[c] | | | | | | | |
| $\beta_X$ | 21.67 | 2.28 | 1.02 | 0.08 | 0.35 | 0.09 | 0.44 | 0.11 | 0.05 | 0.08 |
| $\beta_R$ | 8.59 | 1.25 | 0.67 | 0.61 | 0.32 | 0.92 | 0.23 | 0.19 | 0.04 | 0.11 |
| $\beta_T$ | 105.49 | 16.47 | 97.13 | 8.04 | 12.46 | 3.03 | 2.72 | 1.84 | 0.44 | 1.19 |
| Earth | 4.95 | 0.47 | 0.30 | 0.36 | 0.09 | 0.16 | 0.14 | 0.05 | 0.02 | 0.05 |
| Moon | 8.43 | 1.56 | 0.04 | 1.17 | 0.03 | 0.41 | 0.01 | 0.25 | 0.01 | 0.49 |
| Doppler | 69.79 | 11.46 | 55.45 | 6.81 | 7.37 | 2.05 | 3.00 | 1.84 | 0.48 | 1.23 |
| Spin | 6.00 | 0.75 | 0.23 | 0.37 | 0.19 | 0.30 | 0.17 | 0.25 | 0.03 | 0.15 |
| Other | 63.04 | 5.52 | 0.91 | 0.85 | 0.54 | 0.65 | 0.61 | 0.60 | 0.46 | 0.68 |
| TOTAL[d] | 143.70 | 21.05 | 111.86 | 10.67 | 14.49 | 3.86 | 4.13 | 2.70 | 0.80 | 1.92 |
| | | | | | | | | | | |
| | | | Channel 53A After Correction[e] | | | | | | | |
| $\beta_X$ | 2.86 | 0.30 | 0.13 | 0.01 | 0.05 | 0.01 | 0.06 | 0.01 | 0.01 | 0.01 |
| $\beta_R$ | 4.52 | 0.66 | 0.35 | 0.32 | 0.17 | 0.48 | 0.12 | 0.10 | 0.02 | 0.06 |
| $\beta_T$ | 15.61 | 2.44 | 14.38 | 1.19 | 1.84 | 0.45 | 0.40 | 0.27 | 0.06 | 0.18 |
| Earth | 7.42 | 0.71 | 0.44 | 0.54 | 0.13 | 0.24 | 0.20 | 0.08 | 0.04 | 0.07 |
| Moon | 1.05 | 0.19 | 0.00 | 0.15 | 0.00 | 0.05 | 0.00 | 0.03 | 0.00 | 0.06 |
| Doppler | 1.29 | 0.21 | 0.89 | 0.15 | 0.12 | 0.05 | 0.05 | 0.03 | 0.01 | 0.02 |
| Spin | 5.88 | 0.73 | 0.23 | 0.36 | 0.18 | 0.30 | 0.17 | 0.25 | 0.03 | 0.15 |
| Other | 11.30 | 0.61 | 0.10 | 0.15 | 0.10 | 0.14 | 0.10 | 0.13 | 0.10 | 0.15 |
| TOTAL[d] | 22.19 | 2.82 | 14.42 | 1.42 | 1.87 | 0.78 | 0.51 | 0.41 | 0.13 | 0.30 |

[a] The peak-to-peak amplitude in the map (after the best-fit dipole has been removed).

[b] The pixel-to-pixel standard deviation (after the best-fit dipole has been removed).

[c] Uncertainties refer to antenna temperature of the individual channel or $(A + B)/2$ "sum" maps.

[d] The Total refers to the sum, in quadrature, of the individual effects.

[e] After correction $\Delta T$ terms are antenna temperature 95% confidence level upper limits.

Table 14: Quadrature Sum of Systematic Effects for Channel 53B

| Effect | P-P[a] | RMS[b] | $\Delta T_1$ | $\Delta T_2$ | $\Delta T_3$ | $\Delta T_4$ | $\Delta T_5$ | $\Delta T_6$ | $\Delta T_7$ | $\Delta T_8$ |
|---|---|---|---|---|---|---|---|---|---|---|
| | ($\mu$K) | ($\mu$K) | ($\mu$K) | ($\mu$K) | ($\mu$K) | ($\mu$K) | ($\mu$K) | ($\mu$K) | ($\mu$K) | ($\mu$K) |
| | | | | | | | | | | |
| | | | Channel 53B Before Correction[c] | | | | | | | |
| $\beta_X$ | 1.21 | 0.12 | 0.03 | 0.01 | 0.01 | 0.01 | 0.02 | 0.00 | 0.00 | 0.00 |
| $\beta_R$ | 28.20 | 4.09 | 2.19 | 2.00 | 1.06 | 3.00 | 0.74 | 0.64 | 0.12 | 0.37 |
| $\beta_T$ | 22.89 | 3.57 | 21.08 | 1.74 | 2.70 | 0.65 | 0.59 | 0.40 | 0.09 | 0.25 |
| Earth | 4.95 | 0.49 | 0.31 | 0.38 | 0.09 | 0.17 | 0.14 | 0.06 | 0.02 | 0.05 |
| Moon | 8.87 | 1.62 | 0.04 | 1.21 | 0.03 | 0.43 | 0.01 | 0.24 | 0.01 | 0.50 |
| Doppler | 70.23 | 11.44 | 55.43 | 6.81 | 7.35 | 2.04 | 2.99 | 1.86 | 0.45 | 1.22 |
| Spin | 6.00 | 0.75 | 0.23 | 0.37 | 0.19 | 0.30 | 0.17 | 0.25 | 0.03 | 0.15 |
| Other | 62.02 | 5.63 | 0.52 | 0.84 | 0.42 | 0.64 | 0.42 | 0.59 | 0.47 | 0.68 |
| TOTAL[d] | 101.18 | 13.98 | 59.35 | 7.47 | 7.92 | 3.78 | 3.17 | 2.13 | 0.67 | 1.55 |
| | | | | | | | | | | |
| | | | Channel 53B After Correction[e] | | | | | | | |
| $\beta_X$ | 0.29 | 0.03 | 0.01 | 0.00 | 0.00 | 0.00 | 0.00 | 0.00 | 0.00 | 0.00 |
| $\beta_R$ | 7.67 | 1.11 | 0.60 | 0.55 | 0.29 | 0.82 | 0.20 | 0.17 | 0.03 | 0.10 |
| $\beta_T$ | 8.42 | 1.31 | 7.76 | 0.64 | 0.99 | 0.24 | 0.22 | 0.15 | 0.03 | 0.09 |
| Earth | 8.03 | 0.80 | 0.51 | 0.62 | 0.14 | 0.27 | 0.23 | 0.09 | 0.03 | 0.07 |
| Moon | 1.10 | 0.20 | 0.00 | 0.15 | 0.00 | 0.05 | 0.00 | 0.03 | 0.00 | 0.06 |
| Doppler | 1.30 | 0.21 | 0.89 | 0.15 | 0.12 | 0.05 | 0.05 | 0.04 | 0.01 | 0.02 |
| Spin | 8.93 | 1.12 | 0.34 | 0.55 | 0.28 | 0.45 | 0.26 | 0.38 | 0.04 | 0.22 |
| Other | 11.30 | 0.61 | 0.10 | 0.15 | 0.09 | 0.14 | 0.09 | 0.13 | 0.10 | 0.15 |
| TOTAL[d] | 20.12 | 2.30 | 7.86 | 1.21 | 1.09 | 1.01 | 0.47 | 0.47 | 0.12 | 0.32 |

[a]The peak-to-peak amplitude in the map (after the best-fit dipole has been removed).

[b]The pixel-to-pixel standard deviation (after the best-fit dipole has been removed).

[c]Uncertainties refer to antenna temperature of the individual channel or $(A + B)/2$ "sum" maps.

[d]The Total refers to the sum, in quadrature, of the individual effects.

[e]After correction $\Delta T$ terms are antenna temperature 95% confidence level upper limits.

Table 15: Quadrature Sum of Systematic Effects for Channel 90A

| Effect | P-P[a] ($\mu$K) | RMS[b] ($\mu$K) | $\Delta T_1$ ($\mu$K) | $\Delta T_2$ ($\mu$K) | $\Delta T_3$ ($\mu$K) | $\Delta T_4$ ($\mu$K) | $\Delta T_5$ ($\mu$K) | $\Delta T_6$ ($\mu$K) | $\Delta T_7$ ($\mu$K) | $\Delta T_8$ ($\mu$K) |
|---|---|---|---|---|---|---|---|---|---|---|
| | | | | | Channel 90A Before Correction[c] | | | | | |
| $\beta_X$ | 3.92 | 0.29 | 0.22 | 0.01 | 0.05 | 0.02 | 0.09 | 0.02 | 0.01 | 0.01 |
| $\beta_R$ | 91.13 | 13.34 | 7.05 | 6.52 | 3.45 | 9.84 | 2.42 | 1.97 | 0.42 | 1.21 |
| $\beta_T$ | 36.45 | 5.77 | 33.87 | 2.81 | 4.37 | 1.06 | 0.96 | 0.63 | 0.16 | 0.43 |
| Earth | 3.58 | 0.41 | 0.29 | 0.31 | 0.07 | 0.14 | 0.12 | 0.07 | 0.02 | 0.03 |
| Moon | 10.03 | 1.88 | 0.08 | 1.41 | 0.04 | 0.51 | 0.02 | 0.32 | 0.01 | 0.59 |
| Doppler | 61.79 | 10.07 | 48.56 | 5.98 | 6.49 | 1.82 | 2.63 | 1.56 | 0.45 | 1.11 |
| Spin | 6.00 | 0.75 | 0.23 | 0.37 | 0.19 | 0.30 | 0.17 | 0.25 | 0.03 | 0.15 |
| Other | 73.32 | 7.85 | 1.59 | 0.77 | 0.63 | 0.61 | 0.45 | 0.55 | 0.43 | 0.62 |
| TOTAL[d] | 137.81 | 19.46 | 59.65 | 9.43 | 8.57 | 10.10 | 3.74 | 2.69 | 0.77 | 1.91 |
| | | | | | Channel 90A After Correction[e] | | | | | |
| $\beta_X$ | 1.03 | 0.08 | 0.06 | 0.00 | 0.01 | 0.01 | 0.02 | 0.01 | 0.00 | 0.00 |
| $\beta_R$ | 13.30 | 1.95 | 1.03 | 0.95 | 0.50 | 1.44 | 0.35 | 0.29 | 0.06 | 0.18 |
| $\beta_T$ | 13.41 | 2.12 | 12.47 | 1.03 | 1.61 | 0.39 | 0.35 | 0.23 | 0.06 | 0.16 |
| Earth | 9.05 | 1.03 | 0.73 | 0.78 | 0.17 | 0.36 | 0.30 | 0.17 | 0.04 | 0.07 |
| Moon | 1.64 | 0.31 | 0.01 | 0.23 | 0.01 | 0.08 | 0.00 | 0.05 | 0.00 | 0.10 |
| Doppler | 2.54 | 0.42 | 1.94 | 0.26 | 0.26 | 0.08 | 0.11 | 0.06 | 0.02 | 0.05 |
| Spin | 11.45 | 1.43 | 0.44 | 0.70 | 0.36 | 0.58 | 0.33 | 0.48 | 0.05 | 0.28 |
| Other | 8.19 | 0.61 | 0.24 | 0.09 | 0.10 | 0.09 | 0.09 | 0.08 | 0.06 | 0.09 |
| TOTAL[d] | 25.44 | 3.47 | 12.69 | 1.79 | 1.75 | 1.64 | 0.68 | 0.64 | 0.13 | 0.40 |

[a]The peak-to-peak amplitude in the map (after the best-fit dipole has been removed).

[b]The pixel-to-pixel standard deviation (after the best-fit dipole has been removed).

[c]Uncertainties refer to antenna temperature of the individual channel or $(A + B)/2$ "sum" maps.

[d]The Total refers to the sum, in quadrature, of the individual effects.

[e]After correction $\Delta T$ terms are antenna temperature 95% confidence level upper limits.

Table 16: Quadrature Sum of Systematic Effects for Channel 90B

| Effect | P-P[a] | RMS[b] | $\Delta T_1$ | $\Delta T_2$ | $\Delta T_3$ | $\Delta T_4$ | $\Delta T_5$ | $\Delta T_6$ | $\Delta T_7$ | $\Delta T_8$ |
|--------|--------|--------|--------------|--------------|--------------|--------------|--------------|--------------|--------------|--------------|
| | ($\mu$K) | ($\mu$K) | ($\mu$K) | ($\mu$K) | ($\mu$K) | ($\mu$K) | ($\mu$K) | ($\mu$K) | ($\mu$K) | ($\mu$K) |
| | | | | | | | | | | |
| Channel 90B Before Correction[c] | | | | | | | | | | |
| $\beta_X$ | 0.12 | 0.01 | 0.01 | 0.00 | 0.00 | 0.00 | 0.00 | 0.00 | 0.00 | 0.00 |
| $\beta_R$ | 10.18 | 1.50 | 0.79 | 0.73 | 0.39 | 1.10 | 0.27 | 0.22 | 0.05 | 0.14 |
| $\beta_T$ | 19.91 | 3.14 | 18.42 | 1.53 | 2.38 | 0.58 | 0.52 | 0.34 | 0.09 | 0.23 |
| Earth | 3.97 | 0.41 | 0.30 | 0.31 | 0.07 | 0.14 | 0.11 | 0.05 | 0.02 | 0.04 |
| Moon | 11.17 | 2.07 | 0.08 | 1.54 | 0.04 | 0.56 | 0.02 | 0.32 | 0.01 | 0.64 |
| Doppler | 62.48 | 10.08 | 48.55 | 5.98 | 6.50 | 1.82 | 2.64 | 1.56 | 0.46 | 1.11 |
| Spin | 6.00 | 0.75 | 0.23 | 0.37 | 0.19 | 0.30 | 0.17 | 0.25 | 0.03 | 0.15 |
| Other | 67.54 | 6.31 | 1.82 | 0.78 | 0.69 | 0.61 | 0.47 | 0.55 | 0.43 | 0.63 |
| TOTAL[d] | 95.61 | 12.59 | 51.97 | 6.47 | 6.97 | 2.38 | 2.75 | 1.75 | 0.64 | 1.46 |
| | | | | | | | | | | |
| Channel 90B After Correction[e] | | | | | | | | | | |
| $\beta_X$ | 0.54 | 0.04 | 0.02 | 0.00 | 0.01 | 0.00 | 0.01 | 0.00 | 0.00 | 0.00 |
| $\beta_R$ | 4.07 | 0.60 | 0.32 | 0.29 | 0.15 | 0.44 | 0.11 | 0.09 | 0.02 | 0.05 |
| $\beta_T$ | 9.24 | 1.46 | 8.55 | 0.71 | 1.10 | 0.27 | 0.24 | 0.16 | 0.04 | 0.11 |
| Earth | 7.32 | 0.76 | 0.56 | 0.58 | 0.13 | 0.26 | 0.21 | 0.09 | 0.03 | 0.08 |
| Moon | 1.81 | 0.34 | 0.01 | 0.25 | 0.01 | 0.09 | 0.00 | 0.05 | 0.00 | 0.10 |
| Doppler | 1.84 | 0.30 | 1.36 | 0.19 | 0.18 | 0.06 | 0.08 | 0.05 | 0.01 | 0.03 |
| Spin | 5.70 | 0.71 | 0.22 | 0.35 | 0.18 | 0.29 | 0.16 | 0.24 | 0.02 | 0.14 |
| Other | 8.45 | 0.62 | 0.21 | 0.09 | 0.09 | 0.09 | 0.09 | 0.08 | 0.06 | 0.09 |
| TOTAL[d] | 16.32 | 2.04 | 8.68 | 1.07 | 1.15 | 0.66 | 0.39 | 0.33 | 0.09 | 0.25 |

[a]The peak-to-peak amplitude in the map (after the best-fit dipole has been removed).

[b]The pixel-to-pixel standard deviation (after the best-fit dipole has been removed).

[c]Uncertainties refer to antenna temperature of the individual channel or $(A + B)/2$ "sum" maps.

[d]The Total refers to the sum, in quadrature, of the individual effects.

[e]After correction $\Delta T$ terms are antenna temperature 95% confidence level upper limits.

Table 17: Quadrature Sum of Systematic Effects for Channel 31A Quadrupole Components

| Effect | $Q_1$ | $Q_2$ | $Q_3$ | $Q_4$ | $Q_5$ | $Q_{rms}$ |
|---|---|---|---|---|---|---|
| 31A Before Correction[a] | | | | | | |
| $\beta_X$ | 0.03 | 0.24 | 0.00 | 0.08 | 0.14 | 0.15 |
| $\beta_R$ | 2.07 | 8.69 | 0.16 | 4.05 | 5.06 | 5.67 |
| $\beta_T$ | 0.95 | 1.41 | 0.80 | 0.51 | 2.38 | 1.57 |
| Earth | 0.29 | 0.15 | 0.12 | 0.17 | 0.41 | 0.28 |
| Moon | 0.05 | 0.14 | 1.93 | 1.46 | 0.16 | 1.25 |
| Doppler | 2.75 | 7.64 | 6.19 | 0.89 | 6.20 | 6.14 |
| Spin | 0.34 | 0.24 | 0.26 | 0.49 | 0.24 | 0.37 |
| Other | 0.25 | 0.82 | 0.53 | 0.71 | 0.86 | 0.78 |
| TOTAL[b] | 3.61 | 11.69 | 6.56 | 4.52 | 8.41 | 8.65 |
| | | | | | | |
| 31A After Correction[c] | | | | | | |
| $\beta_X$ | 0.01 | 0.04 | 0.00 | 0.01 | 0.02 | 0.02 |
| $\beta_R$ | 0.87 | 3.63 | 0.07 | 1.69 | 2.11 | 2.37 |
| $\beta_T$ | 0.96 | 1.43 | 0.81 | 0.52 | 2.41 | 1.59 |
| Earth | 0.52 | 0.26 | 0.22 | 0.29 | 0.73 | 0.50 |
| Moon | 0.01 | 0.01 | 0.21 | 0.16 | 0.02 | 0.14 |
| Doppler | 0.16 | 0.38 | 0.32 | 0.05 | 0.31 | 0.31 |
| Spin | 0.75 | 0.52 | 0.59 | 1.08 | 0.53 | 0.81 |
| Other | 0.03 | 0.11 | 0.11 | 0.13 | 0.10 | 0.12 |
| TOTAL | 1.59 | 3.97 | 1.10 | 2.11 | 3.35 | 3.03 |

[a]All terms refer to best estimate of the uncorrected effect, in $\mu$K antenna temperature

[b]The Total refers to the sum, in quadrature, of the items above

[c]95% confidence level upper limits, in $\mu$K antenna temperature

Table 18: Quadrature Sum of Systematic Effects for Channel 31B Quadrupole Components

| Effect | $Q_1$ | $Q_2$ | $Q_3$ | $Q_4$ | $Q_5$ | $Q_{rms}$ |
|---|---|---|---|---|---|---|
| **31B Before Correction**[a] | | | | | | |
| $\beta_X$ | 0.06 | 0.12 | 0.02 | 0.00 | 0.08 | 0.08 |
| $\beta_R$ | 7.42 | 10.22 | 1.11 | 0.54 | 7.78 | 7.44 |
| $\beta_T$ | 0.12 | 0.55 | 0.33 | 0.08 | 0.79 | 0.53 |
| Earth | 0.38 | 0.18 | 0.13 | 0.19 | 0.44 | 0.32 |
| Moon | 0.09 | 0.06 | 1.95 | 1.45 | 0.18 | 1.26 |
| Doppler | 1.72 | 10.67 | 9.09 | 2.56 | 3.91 | 7.67 |
| Spin | 0.34 | 0.24 | 0.26 | 0.49 | 0.24 | 0.37 |
| Other | 0.43 | 0.84 | 0.39 | 0.89 | 0.86 | 0.82 |
| TOTAL[b] | 7.64 | 14.82 | 9.38 | 3.16 | 8.80 | 10.82 |
| | | | | | | |
| **31B After Correction**[c] | | | | | | |
| $\beta_X$ | 0.07 | 0.13 | 0.02 | 0.00 | 0.08 | 0.09 |
| $\beta_R$ | 3.83 | 5.28 | 0.57 | 0.28 | 4.01 | 3.84 |
| $\beta_T$ | 0.38 | 1.72 | 1.03 | 0.24 | 2.47 | 1.66 |
| Earth | 0.71 | 0.34 | 0.24 | 0.35 | 0.81 | 0.59 |
| Moon | 0.01 | 0.01 | 0.21 | 0.15 | 0.02 | 0.13 |
| Doppler | 0.09 | 0.49 | 0.45 | 0.13 | 0.19 | 0.36 |
| Spin | 1.61 | 1.12 | 1.26 | 2.32 | 1.13 | 1.75 |
| Other | 0.08 | 0.15 | 0.10 | 0.16 | 0.12 | 0.14 |
| TOTAL[b] | 4.23 | 5.70 | 1.81 | 2.39 | 4.92 | 4.59 |

[a]All terms refer to best estimate of the uncorrected effect, in $\mu$K antenna temperature

[b]The Total refers to the sum, in quadrature, of the items above

[c]95% confidence level upper limits, in $\mu$K antenna temperature

Table 19: Quadrature Sum of Systematic Effects for Channel 53A Quadrupole Components

| Effect | $Q_1$ | $Q_2$ | $Q_3$ | $Q_4$ | $Q_5$ | $Q_{rms}$ |
|---|---|---|---|---|---|---|
| **53A Before Correction**[a] | | | | | | |
| $\beta_X$ | 0.05 | 0.02 | 0.01 | 0.07 | 0.13 | 0.08 |
| $\beta_R$ | 0.13 | 0.68 | 0.77 | 0.51 | 0.28 | 0.61 |
| $\beta_T$ | 5.21 | 6.63 | 3.11 | 2.26 | 12.78 | 8.04 |
| Earth | 0.25 | 0.29 | 0.03 | 0.20 | 0.57 | 0.36 |
| Moon | 0.05 | 0.13 | 1.81 | 1.35 | 0.15 | 1.17 |
| Doppler | 7.14 | 5.29 | 8.07 | 3.00 | 5.79 | 6.81 |
| Spin | 0.34 | 0.24 | 0.26 | 0.49 | 0.24 | 0.37 |
| Other | 0.28 | 0.88 | 0.51 | 0.78 | 0.99 | 0.85 |
| TOTAL[b] | 8.85 | 8.56 | 8.89 | 4.14 | 14.08 | 10.67 |
| | | | | | | |
| **53A After Correction**[c] | | | | | | |
| $\beta_X$ | 0.01 | 0.00 | 0.00 | 0.01 | 0.02 | 0.01 |
| $\beta_R$ | 0.07 | 0.36 | 0.40 | 0.27 | 0.14 | 0.32 |
| $\beta_T$ | 0.77 | 0.98 | 0.46 | 0.33 | 1.89 | 1.19 |
| Earth | 0.38 | 0.43 | 0.04 | 0.30 | 0.85 | 0.54 |
| Moon | 0.01 | 0.02 | 0.22 | 0.17 | 0.02 | 0.15 |
| Doppler | 0.18 | 0.10 | 0.18 | 0.06 | 0.10 | 0.15 |
| Spin | 0.33 | 0.23 | 0.26 | 0.48 | 0.23 | 0.36 |
| Other | 0.05 | 0.15 | 0.11 | 0.18 | 0.13 | 0.15 |
| TOTAL[b] | 0.94 | 1.17 | 0.73 | 0.75 | 2.10 | 1.42 |

[a]All terms refer to best estimate of the uncorrected effect, in $\mu$K antenna temperature

[b]The Total refers to the sum, in quadrature, of the items above

[c]95% confidence level upper limits, in $\mu$K antenna temperature

Table 20: Quadrature Sum of Systematic Effects for Channel 53B Quadrupole Components

| Effect | $Q_1$ | $Q_2$ | $Q_3$ | $Q_4$ | $Q_5$ | $Q_{rms}$ |
|---|---|---|---|---|---|---|
| **53B Before Correction**[a] | | | | | | |
| $\beta_X$ | 0.00 | 0.00 | 0.00 | 0.00 | 0.01 | 0.01 |
| $\beta_R$ | 0.43 | 2.24 | 2.52 | 1.66 | 0.90 | 2.00 |
| $\beta_T$ | 1.13 | 1.44 | 0.68 | 0.49 | 2.77 | 1.74 |
| Earth | 0.28 | 0.31 | 0.02 | 0.20 | 0.59 | 0.38 |
| Moon | 0.06 | 0.13 | 1.86 | 1.39 | 0.15 | 1.21 |
| Doppler | 7.14 | 5.28 | 8.08 | 3.01 | 5.78 | 6.81 |
| Spin | 0.34 | 0.24 | 0.26 | 0.49 | 0.24 | 0.37 |
| Other | 0.27 | 0.88 | 0.51 | 0.79 | 0.97 | 0.84 |
| TOTAL[b] | 7.26 | 5.99 | 8.71 | 3.86 | 6.58 | 7.47 |
| | | | | | | |
| **53B After Correction**[c] | | | | | | |
| $\beta_X$ | 0.00 | 0.00 | 0.00 | 0.00 | 0.00 | 0.00 |
| $\beta_R$ | 0.12 | 0.61 | 0.68 | 0.45 | 0.25 | 0.55 |
| $\beta_T$ | 0.42 | 0.53 | 0.25 | 0.18 | 1.02 | 0.64 |
| Earth | 0.45 | 0.49 | 0.04 | 0.33 | 0.96 | 0.62 |
| Moon | 0.01 | 0.02 | 0.23 | 0.17 | 0.02 | 0.15 |
| Doppler | 0.18 | 0.10 | 0.18 | 0.06 | 0.10 | 0.15 |
| Spin | 0.50 | 0.35 | 0.39 | 0.73 | 0.35 | 0.55 |
| Other | 0.05 | 0.15 | 0.12 | 0.18 | 0.13 | 0.16 |
| TOTAL[b] | 0.82 | 1.02 | 0.89 | 0.97 | 1.48 | 1.21 |

[a]All terms refer to best estimate of the uncorrected effect, in $\mu$K antenna temperature

[b]The Total refers to the sum, in quadrature, of the items above

[c]95% confidence level upper limits, in $\mu$K antenna temperature

Table 21: Quadrature Sum of Systematic Effects for Channel 90A Quadrupole Components

| Effect | $Q_1$ | $Q_2$ | $Q_3$ | $Q_4$ | $Q_5$ | $Q_{rms}$ |
|---|---|---|---|---|---|---|
| **90A Before Correction**[a] | | | | | | |
| $\beta_X$ | 0.02 | 0.01 | 0.00 | 0.01 | 0.02 | 0.01 |
| $\beta_R$ | 1.50 | 7.21 | 8.27 | 5.39 | 2.88 | 6.52 |
| $\beta_T$ | 1.81 | 2.32 | 1.09 | 0.78 | 4.47 | 2.81 |
| Earth | 0.23 | 0.25 | 0.03 | 0.16 | 0.48 | 0.31 |
| Moon | 0.06 | 0.15 | 2.18 | 1.62 | 0.18 | 1.41 |
| Doppler | 6.26 | 4.64 | 7.08 | 2.66 | 5.09 | 5.98 |
| Spin | 0.34 | 0.24 | 0.26 | 0.49 | 0.24 | 0.37 |
| Other | 0.24 | 0.79 | 0.56 | 0.71 | 0.88 | 0.77 |
| TOTAL[b] | 6.71 | 8.92 | 11.17 | 6.34 | 7.43 | 9.43 |
| | | | | | | |
| **90A After Correction**[c] | | | | | | |
| $\beta_X$ | 0.01 | 0.00 | 0.00 | 0.00 | 0.00 | 0.00 |
| $\beta_R$ | 0.22 | 1.05 | 1.21 | 0.79 | 0.42 | 0.95 |
| $\beta_T$ | 0.67 | 0.85 | 0.40 | 0.29 | 1.64 | 1.03 |
| Earth | 0.59 | 0.64 | 0.07 | 0.39 | 1.21 | 0.78 |
| Moon | 0.01 | 0.03 | 0.36 | 0.27 | 0.03 | 0.23 |
| Doppler | 0.28 | 0.19 | 0.31 | 0.11 | 0.21 | 0.26 |
| Spin | 0.65 | 0.45 | 0.50 | 0.93 | 0.45 | 0.70 |
| Other | 0.03 | 0.09 | 0.08 | 0.10 | 0.09 | 0.09 |
| TOTAL[b] | 1.16 | 1.58 | 1.45 | 1.35 | 2.15 | 1.79 |

[a]All terms refer to best estimate of the uncorrected effect, in $\mu$K antenna temperature

[b]The Total refers to the sum, in quadrature, of the items above

[c]95% confidence level upper limits, in $\mu$K antenna temperature

Table 22: Quadrature Sum of Systematic Effects for Channel 90B Quadrupole Components

| Effect | $Q_1$ | $Q_2$ | $Q_3$ | $Q_4$ | $Q_5$ | $Q_{rms}$ |
|---|---|---|---|---|---|---|
| **90B Before Correction**[a] | | | | | | |
| $\beta_X$ | 0.00 | 0.00 | 0.00 | 0.00 | 0.00 | 0.00 |
| $\beta_R$ | 0.17 | 0.81 | 0.92 | 0.61 | 0.32 | 0.73 |
| $\beta_T$ | 0.98 | 1.26 | 0.59 | 0.42 | 2.43 | 1.53 |
| Earth | 0.23 | 0.26 | 0.02 | 0.16 | 0.49 | 0.31 |
| Moon | 0.07 | 0.17 | 2.38 | 1.77 | 0.20 | 1.54 |
| Doppler | 6.27 | 4.64 | 7.07 | 2.66 | 5.09 | 5.98 |
| Spin | 0.34 | 0.24 | 0.26 | 0.49 | 0.24 | 0.37 |
| Other | 0.23 | 0.81 | 0.58 | 0.70 | 0.86 | 0.78 |
| TOTAL[b] | 6.37 | 4.96 | 7.57 | 3.40 | 5.74 | 6.47 |
| | | | | | | |
| **90B After Correction**[c] | | | | | | |
| $\beta_X$ | 0.00 | 0.00 | 0.00 | 0.00 | 0.00 | 0.00 |
| $\beta_R$ | 0.07 | 0.32 | 0.37 | 0.24 | 0.13 | 0.29 |
| $\beta_T$ | 0.46 | 0.58 | 0.27 | 0.20 | 1.13 | 0.71 |
| Earth | 0.42 | 0.47 | 0.04 | 0.30 | 0.90 | 0.58 |
| Moon | 0.01 | 0.03 | 0.39 | 0.29 | 0.03 | 0.25 |
| Doppler | 0.21 | 0.14 | 0.23 | 0.08 | 0.15 | 0.19 |
| Spin | 0.32 | 0.22 | 0.25 | 0.46 | 0.23 | 0.35 |
| Other | 0.03 | 0.09 | 0.08 | 0.10 | 0.08 | 0.09 |
| TOTAL[b] | 0.73 | 0.86 | 0.69 | 0.71 | 1.48 | 1.07 |

[a]All terms refer to best estimate of the uncorrected effect, in $\mu$K antenna temperature

[b]The Total refers to the sum, in quadrature, of the items above

[c]95% confidence level upper limits, in $\mu$K antenna temperature

Table 23: RMS Thermodynamic Temperature Fluctuations

| Data set | $b_{cut}$ | 7° RMS (unit weighting) | | | 10° RMS (unit weighting) | | |
|---|---|---|---|---|---|---|---|
| | | Sum | Diff | Sky | Sum | Diff | Sky |
| | (°) | ($\mu$K) | ($\mu$K) | ($\mu$K) | ($\mu$K) | ($\mu$K) | ($\mu$K) |
| 31 Yr 1 | 10 | 415.1 | 394.8 | 128.1 | 139.0 | 101.9 | 94.5 |
| | 20 | 409.6 | 401.8 | 80.0 | 110.0 | 105.7 | 30.2 |
| | 30 | 417.1 | 411.9 | 65.7 | 106.7 | 107.9 | 0.0 |
| | 40 | 421.8 | 418.8 | 50.0 | 112.7 | 108.2 | 31.6 |
| 31 Yr 2 | 10 | 460.2 | 439.6 | 136.4 | 143.9 | 113.0 | 89.1 |
| | 20 | 464.6 | 449.8 | 116.5 | 124.1 | 115.6 | 45.3 |
| | 30 | 474.3 | 462.5 | 104.9 | 123.5 | 116.3 | 41.6 |
| | 40 | 488.6 | 474.4 | 116.9 | 121.6 | 124.7 | 0.0 |
| 31 2 Yr | 10 | 310.5 | 290.8 | $109.0 \pm 27.5$ | 117.5 | 76.0 | $89.7 \pm 9.5$ |
| | 20 | 303.6 | 295.7 | $69.2 \pm 29.5$ | 85.3 | 77.6 | $35.5 \pm 11.2$ |
| | 30 | 306.8 | 305.0 | $34.0 \pm 31.3$ | 80.9 | 78.1 | $21.1 \pm 13.2$ |
| | 40 | 314.0 | 310.0 | $49.8 \pm 34.9$ | 83.7 | 78.8 | $27.9 \pm 15.5$ |
| 53 Yr 1 | 10 | 160.7 | 152.0 | 52.3 | 56.0 | 38.2 | 41.0 |
| | 20 | 158.2 | 151.6 | 45.2 | 49.3 | 37.7 | 31.7 |
| | 30 | 159.0 | 151.9 | 47.1 | 48.9 | 38.4 | 30.3 |
| | 40 | 159.9 | 153.3 | 45.4 | 50.3 | 39.4 | 31.2 |
| 53 Yr 2 | 10 | 162.7 | 149.8 | 63.5 | 57.5 | 38.6 | 42.6 |
| | 20 | 160.6 | 149.3 | 59.0 | 50.2 | 38.6 | 32.2 |
| | 30 | 158.6 | 149.1 | 54.2 | 49.9 | 39.4 | 30.6 |
| | 40 | 156.9 | 147.5 | 53.4 | 47.4 | 38.9 | 27.1 |
| 53 2 Yr | 10 | 118.9 | 105.9 | $54.2 \pm 6.4$ | 50.2 | 26.7 | $42.5 \pm 2.2$ |
| | 20 | 115.2 | 106.3 | $44.4 \pm 7.5$ | 42.1 | 26.9 | $32.4 \pm 2.5$ |
| | 30 | 114.6 | 105.8 | $43.9 \pm 9.0$ | 40.9 | 27.2 | $30.5 \pm 2.7$ |
| | 40 | 115.1 | 103.8 | $49.7 \pm 10.9$ | 40.6 | 27.2 | $30.2 \pm 3.2$ |
| 90 Yr 1 | 10 | 243.4 | 243.0 | 13.9 | 70.3 | 61.4 | 34.3 |
| | 20 | 242.5 | 243.7 | 0.0 | 65.7 | 64.0 | 14.9 |
| | 30 | 243.1 | 244.1 | 0.0 | 67.2 | 62.4 | 24.9 |
| | 40 | 241.8 | 245.2 | 0.0 | 66.6 | 61.6 | 25.3 |
| 90 Yr 2 | 10 | 247.7 | 244.7 | 38.5 | 75.1 | 64.4 | 38.5 |
| | 20 | 245.1 | 244.6 | 15.7 | 68.7 | 64.5 | 23.8 |
| | 30 | 243.0 | 243.9 | 0.0 | 69.7 | 61.2 | 33.3 |
| | 40 | 242.0 | 243.9 | 0.0 | 67.8 | 62.3 | 26.8 |
| 90 2 Yr | 10 | 175.5 | 174.4 | $20.0 \pm 10.1$ | 59.0 | 44.5 | $38.7 \pm 3.2$ |
| | 20 | 173.3 | 173.8 | $0.0 \pm 11.9$ | 52.8 | 46.1 | $25.7 \pm 3.8$ |
| | 30 | 174.5 | 171.8 | $30.2 \pm 14.3$ | 54.4 | 41.8 | $34.8 \pm 4.1$ |
| | 40 | 172.8 | 173.1 | $0.0 \pm 17.4$ | 54.4 | 42.6 | $33.9 \pm 4.9$ |

Table 24: RMS Thermodynamic Temperature Fluctuations with Galaxy Removed

| Data set | $b_{cut}$ | 7° RMS (unit weighting) | | | 10° RMS (unit weighting) | | |
|---|---|---|---|---|---|---|---|
| | | Sum | Diff | Sky | Sum | Diff | Sky |
| | (°) | ($\mu$K) | ($\mu$K) | ($\mu$K) | ($\mu$K) | ($\mu$K) | ($\mu$K) |
| Comb Yr 1 | 10 | 276.3 | 267.8 | 68.0 | 73.6 | 67.2 | 30.0 |
| | 20 | 277.8 | 271.0 | 61.1 | 73.5 | 67.9 | 28.0 |
| | 30 | 283.6 | 273.5 | 74.9 | 76.7 | 70.4 | 30.4 |
| | 40 | 285.2 | 277.9 | 64.5 | 79.3 | 71.5 | 34.4 |
| Comb Yr 2 | 10 | 299.2 | 283.1 | 96.7 | 81.1 | 72.1 | 37.1 |
| | 20 | 304.0 | 286.4 | 102.0 | 80.2 | 73.8 | 31.3 |
| | 30 | 307.2 | 291.0 | 98.4 | 81.6 | 74.3 | 33.7 |
| | 40 | 313.6 | 292.4 | 113.4 | 82.3 | 75.4 | 33.0 |
| Comb 2 Yr | 10 | 201.9 | 194.0 | $55.7 \pm 15.4$ | 58.5 | 49.6 | $31.1 \pm 5.3$ |
| | 20 | 203.8 | 197.0 | $52.2 \pm 17.0$ | 57.6 | 51.2 | $26.4 \pm 6.2$ |
| | 30 | 206.0 | 199.1 | $53.2 \pm 18.8$ | 59.4 | 51.9 | $28.9 \pm 7.1$ |
| | 40 | 208.7 | 197.7 | $67.0 \pm 21.8$ | 60.5 | 52.5 | $30.1 \pm 8.4$ |
| Subt Yr 1 | 10 | 227.2 | 223.1 | 42.9 | 63.5 | 55.5 | 30.8 |
| | 20 | 227.7 | 226.3 | 25.0 | 62.5 | 57.0 | 25.8 |
| | 30 | 231.7 | 228.7 | 37.2 | 64.8 | 58.3 | 28.4 |
| | 40 | 232.1 | 230.5 | 27.7 | 67.5 | 58.7 | 33.2 |
| Subt Yr 2 | 10 | 242.3 | 232.5 | 68.2 | 68.9 | 58.4 | 36.5 |
| | 20 | 245.0 | 233.6 | 74.0 | 67.0 | 58.4 | 32.8 |
| | 30 | 247.1 | 236.3 | 72.1 | 67.9 | 57.7 | 35.8 |
| | 40 | 251.4 | 236.1 | 86.1 | 67.8 | 57.7 | 35.6 |
| Subt 2 Yr | 10 | 165.0 | 159.9 | $40.7 \pm 11.9$ | 51.7 | 40.0 | $32.8 \pm 4.0$ |
| | 20 | 165.8 | 161.7 | $36.7 \pm 13.2$ | 50.2 | 41.2 | $28.7 \pm 4.7$ |
| | 30 | 168.0 | 162.4 | $43.3 \pm 14.9$ | 51.3 | 40.1 | $32.1 \pm 5.4$ |
| | 40 | 169.4 | 161.9 | $49.8 \pm 17.4$ | 52.6 | 40.5 | $33.5 \pm 6.4$ |

Table 25: Quadrupole Results[a] After Subtraction-Technique Galaxy Removal

| Quad Comp | Year 1 Sum | Year 1 Diff | Year 2 Sum | Year 2 Diff | 2 Year Sum | 2 Year Diff |
|---|---|---|---|---|---|---|
| $Q_1$ | $9.7 \pm 7.9$ | $-10.1 \pm 7.1$ | $-9.4 \pm 8.4$ | $4.0 \pm 7.7$ | $2.0 \pm 6.5$ | $-4.9 \pm 5.5$ |
| $Q_2$ | $10.1 \pm 6.0$ | $-2.1 \pm 5.9$ | $-2.0 \pm 5.9$ | $-6.6 \pm 5.8$ | $5.0 \pm 4.0$ | $-5.2 \pm 3.9$ |
| $Q_3$ | $14.1 \pm 5.2$ | $-0.1 \pm 5.2$ | $5.3 \pm 5.9$ | $-1.7 \pm 5.9$ | $9.6 \pm 4.0$ | $-0.8 \pm 4.0$ |
| $Q_4$ | $4.9 \pm 7.6$ | $14.3 \pm 6.7$ | $2.2 \pm 8.0$ | $-1.0 \pm 7.1$ | $2.9 \pm 5.9$ | $7.6 \pm 4.7$ |
| $Q_5$ | $4.9 \pm 7.3$ | $-1.5 \pm 6.4$ | $10.3 \pm 7.6$ | $-5.1 \pm 6.7$ | $7.0 \pm 5.9$ | $-2.9 \pm 4.7$ |
| $Q_{rms}$ | 10.6 | 8.7 | 7.5 | 4.8 | 6.9 | 5.5 |
| $\chi^2$ | 12.6 | 6.7 | 4.1 | 2.3 | 9.1 | 5.6 |

[a] Values are in thermodynamic $\mu$K. Uniform weighting of the map pixels was used. The order of the fit was truncated at $\ell = 2$. A Galactic cut of 10° was applied. The kinematic quadrupole correction was applied. A 1.5% correction was applied to restore power lost by beam dilution.

Table 26: Quadrupole Results[a] After Combination-Technique Galaxy Removal

| Quad Comp | Year 1 Sum | Year 1 Diff | Year 2 Sum | Year 2 Diff | 2 Year Sum | 2 Year Diff |
|---|---|---|---|---|---|---|
| $Q_1$ | $10.0 \pm 9.8$ | $-14.6 \pm 9.2$ | $-11.7 \pm 10.7$ | $4.2 \pm 10.1$ | $2.0 \pm 7.4$ | $-7.9 \pm 6.5$ |
| $Q_2$ | $12.6 \pm 6.6$ | $-1.7 \pm 6.5$ | $-1.2 \pm 7.2$ | $-5.4 \pm 7.1$ | $6.9 \pm 5.3$ | $-5.0 \pm 5.2$ |
| $Q_3$ | $13.6 \pm 6.4$ | $1.2 \pm 6.4$ | $9.1 \pm 7.6$ | $-2.3 \pm 7.6$ | $11.0 \pm 5.0$ | $-0.8 \pm 5.0$ |
| $Q_4$ | $0.4 \pm 8.7$ | $12.8 \pm 7.9$ | $3.8 \pm 9.3$ | $-2.3 \pm 8.6$ | $1.0 \pm 7.0$ | $6.6 \pm 6.0$ |
| $Q_5$ | $7.1 \pm 8.6$ | $-2.7 \pm 7.8$ | $14.1 \pm 9.3$ | $-9.5 \pm 8.5$ | $10.1 \pm 6.9$ | $-5.4 \pm 5.8$ |
| $Q_{rms}$ | 11.2 | 9.5 | 10.3 | 6.2 | 8.6 | 6.2 |
| $\chi^2$ | 9.9 | 5.4 | 5.1 | 2.2 | 8.8 | 4.5 |

[a] Values are in thermodynamic $\mu$K. Uniform weighting of the map pixels was used. The order of the fit was truncated at $\ell = 2$. A Galactic cut of 10° was applied. The kinematic quadrupole correction was applied. A 1.5% correction was applied to restore power lost by beam dilution.

## FIGURE CAPTIONS

**Figure 1:** Mollweide projections of the full sky in Galactic coordinates with the Galactic center at the center of the maps and longitude increasing towards the left. The data are smoothed to an effective resolution of 10°. (a) The 31 GHz, 53 GHz, and 90 GHz $(A + B)/2$ (sum) data, and (b) The 31 GHz, 53 GHz, and 90 GHz $(A - B)/2$ (difference) data.

**Figure 2:** Azimuthal equal area projections of the dipole removed 53 GHz $(A + B)/2$ (sum) data and the combination technique Galaxy-reduced data, smoothed to 10°. The North Galactic polar cap is on the left, the South Galactic polar cap is on the right, and the point where they meet is the Galactic center. $\ell = 90°$ is at the bottom of each projected polar cap.

**Figure 3:** The two-point 53 GHz $(A + B)/2 \times$ 90 GHz $(A + B)/2$ cross correlation function with the dipole removed, for $|b| > 20°$ in thermodynamic temperature units. The error bars on the individual points include only instrument noise. *top*: The two-point cross correlation with the quadrupole included. The shaded region is the 68% confidence region expected from a 12.4 $\mu$K, n=1.6 spectrum of fluctuations, including cosmic variance and instrument noise. *bottom*: The two-point cross correlation with the quadrupole excluded. The shaded region is the 68% confidence region expected from a 16.0 $\mu$K, n=1.2 spectrum of fluctuations, including cosmic variance and instrument noise.

**Figure 4:** Likelihood contours as a function of $n$ and $Q_{rms-PS}$ for the data shown in Figure 3. The contours correspond to 68%, 95%, and 99.7% confidence regions. The solid curves are with the quadrupole included, with the peak of the likelihood indicated by +. The dashed curves are with the quadrupole excluded, with the peak of the likelihood indicated by *.

**Figure 5:** Likelihood functions for the observed quadrupole amplitude $Q_{rms}$ in the subtraction technique CMB map. (a) The likelihood of $Q_{rms}$ using only the first year data. The solid curve is for the $(A + B)/2$ sum data and the dashed curve is for the $(A - B)/2$ difference data. (b) The same as (a), but for the second year of data. (c) The same as (a), but for the first two years of data combined.

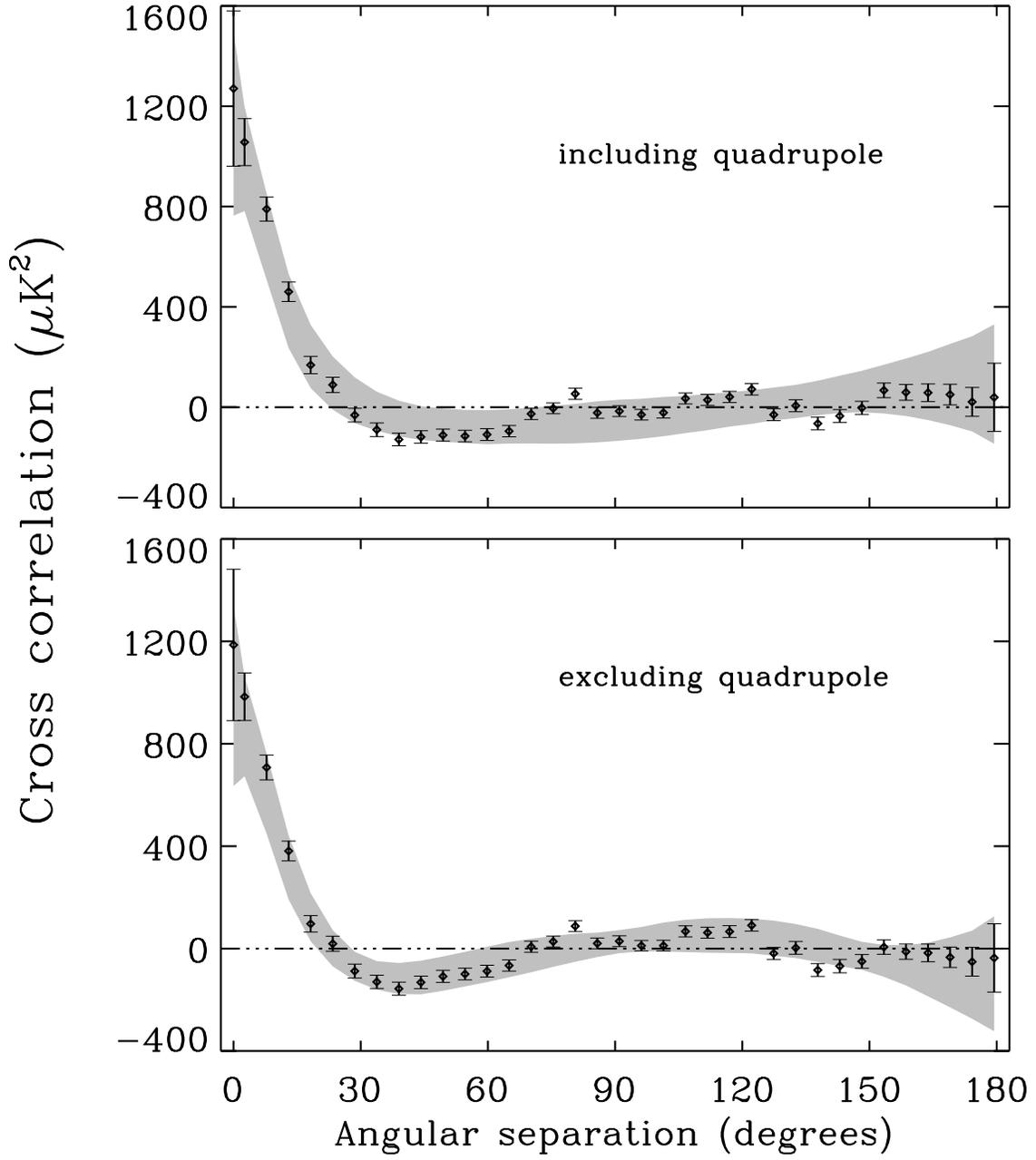

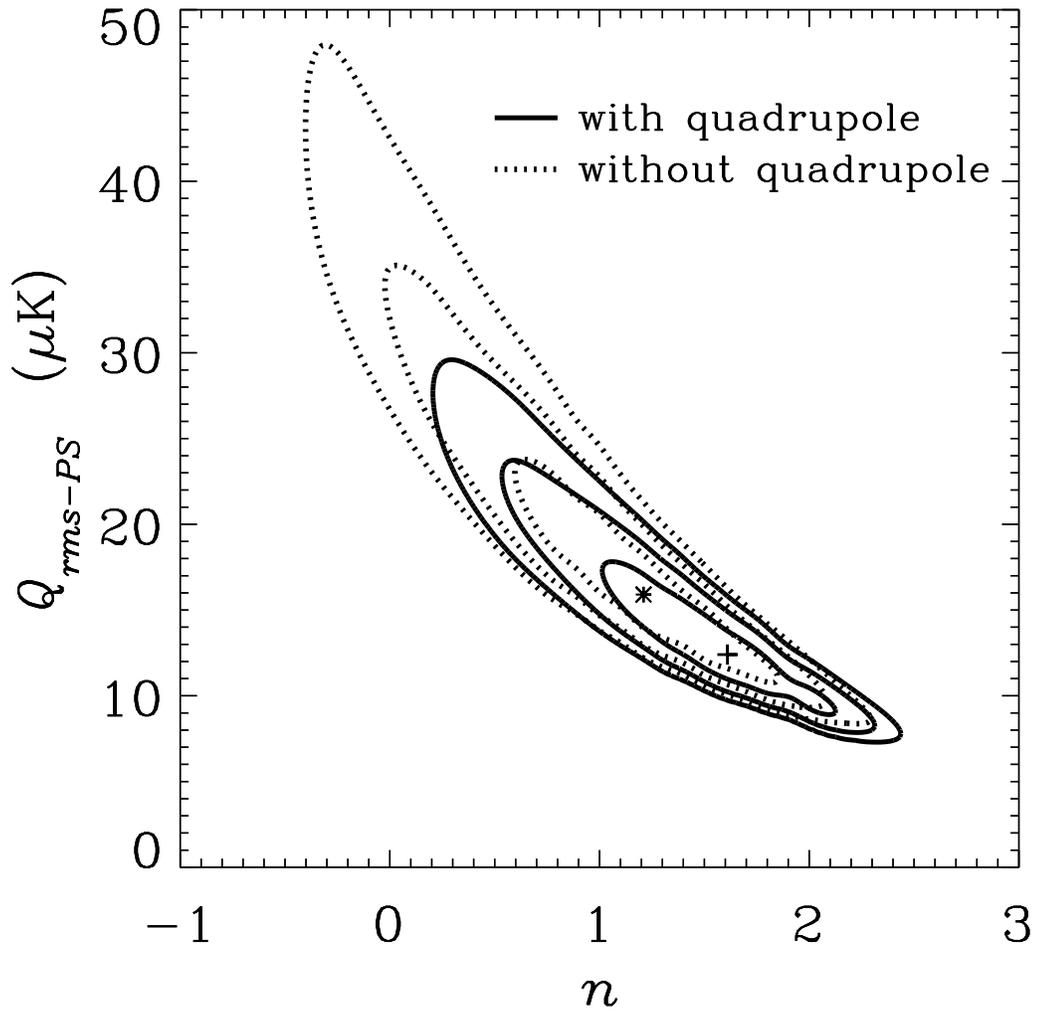

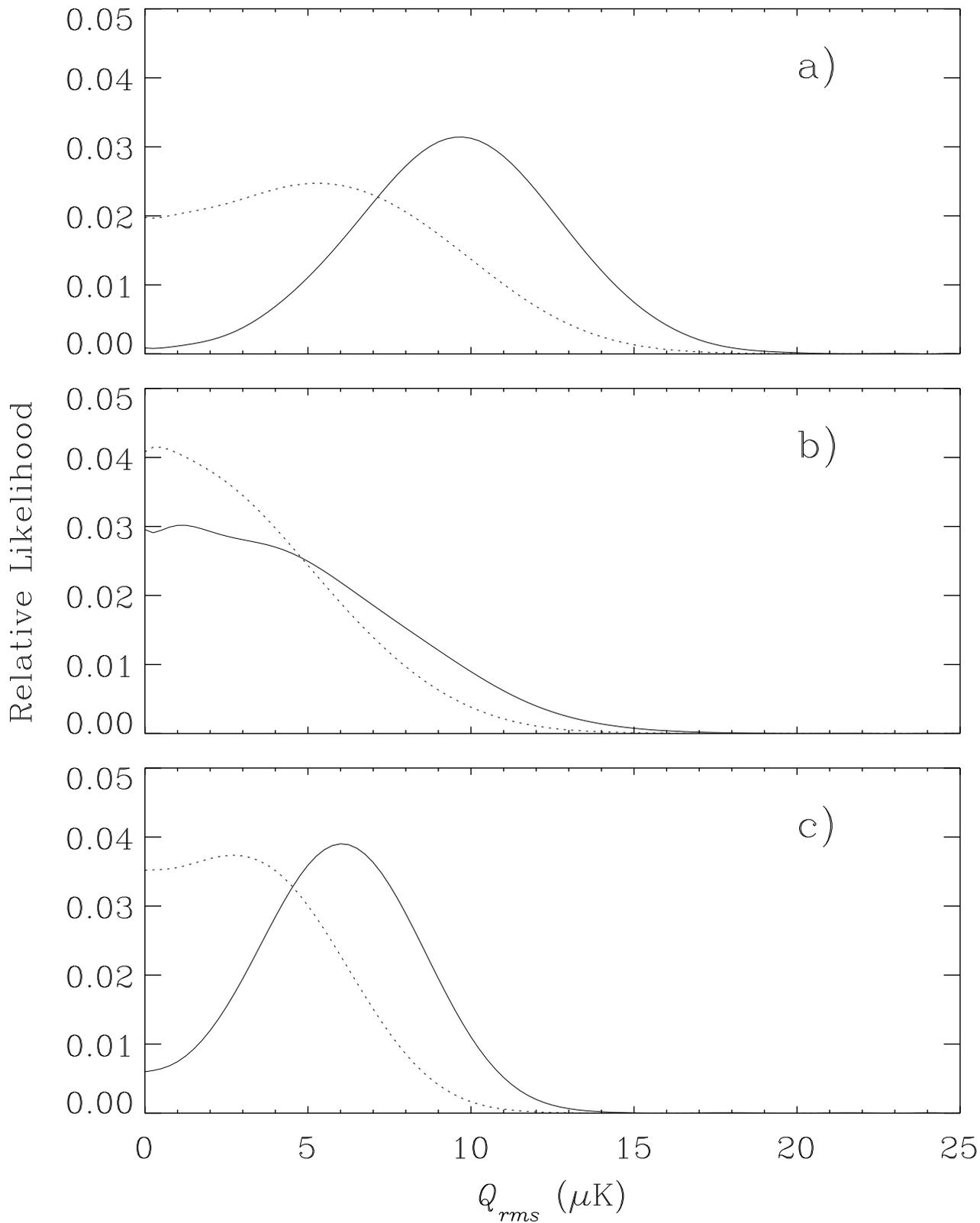